\documentclass[english,prd,superscriptaddress,nofootinbib,preprintnumbers,eqsecnum]{revtex4}
\usepackage{graphicx}
\usepackage{dcolumn}
\usepackage{bm}
\usepackage{latexsym}
\usepackage{amsfonts}
\usepackage{amssymb}
\usepackage{amsmath}
\usepackage{mathrsfs}
\usepackage{color}

\def\be{\begin{equation}}
\def\ee{\end{equation}}
\def\ba{\begin{eqnarray}}
\def\ea{\end{eqnarray}}

\def\D{\bar{D}}

\begin{document}

\title{Effective field theory of modified gravity on the spherically
symmetric background: \\
leading order dynamics and the odd-type perturbations}
\author{Ryotaro Kase}
\affiliation{Department of Physics, Faculty of Science, Tokyo University of Science, 1-3,
Kagurazaka, Shinjuku, Tokyo 162-8601, Japan}
\author{L\'{a}szl\'{o} \'{A}. Gergely}
\affiliation{Department of Physics, Faculty of Science, Tokyo University of Science, 1-3,
Kagurazaka, Shinjuku, Tokyo 162-8601, Japan}
\affiliation{Departments of Theoretical and Experimental Physics, University of Szeged, D%
\'{o}m t\'{e}r 9, 6720 Szeged, Hungary}
\author{Shinji Tsujikawa}
\affiliation{Department of Physics, Faculty of Science, Tokyo University of Science, 1-3,
Kagurazaka, Shinjuku, Tokyo 162-8601, Japan}
\date{\today }

\begin{abstract}
We consider perturbations of a static and spherically symmetric background
endowed with a metric tensor and a scalar field in the framework of the
effective field theory of modified gravity. We employ the previously
developed 2+1+1 canonical formalism of a double Arnowitt-Deser-Misner (ADM)
decomposition of space-time, which singles out both time and radial
directions. Our building block is a general gravitational action that
depends on scalar quantities constructed from the 2+1+1 canonical variables
and the lapse. Variation of the action up to first-order in perturbations
gives rise to three independent background equations of motion, as expected
from spherical symmetry. The dynamical equations of linear perturbations
follow from the second-order Lagrangian after a suitable gauge fixing. We
derive conditions for the avoidance of ghosts and Laplacian instabilities
for the odd-type perturbations. We show that our results not only
incorporate those derived in the most general scalar-tensor theories with
second-order equations of motion (the Horndeski theories) but they can be
applied to more generic theories beyond Horndeski.
\end{abstract}

\maketitle





\section{Introduction}

\label{intro} 

The unexpected discovery of the late-time cosmic acceleration from the
supernovae type-Ia (SN Ia) observations \cite{Riess,Perl} has pushed forward
the idea that the gravitational law may be modified from General Relativity
(GR) at large distances. The recent CMB measurement by Planck \cite{Planck}
combined with data of the WMAP polarization \cite{WMAP} and the SN Ia (from
SNLS \cite{SNLS}) showed that the dark energy equation of state is
constrained to be $w_{\mathrm{DE}}=-1.13^{+0.13}_{-0.14}$ (95\, \%\,CL) for
constant $w_{\mathrm{DE}}$. In GR it is generally difficult to explain $w_{%
\mathrm{DE}}<-1$ unless a ghost mode is introduced, but the modification of
gravity allows a possibility of realizing such an equation of state while
avoiding ghosts and instabilities \cite{review}.

Most of the dark energy models proposed in the literature--such as
quintessence \cite{quin}, k-essence \cite{kes}, $f(R)$ gravity \cite%
{fR1,fRviable}, Brans-Dicke theory \cite{Brans,Yoko}, and Galileons \cite%
{Nicolis,Deffayet1,Deffayet2,Gacosmo}-- belong to the category of the
Horndeski theory \cite{Horndeski,Char,Deffayet11}, i.e., the most general
scalar-tensor theory with second-order equations of motion. In the Horndeski
theory, the conditions for avoiding ghosts and Laplacian instabilities of
scalar and tensor perturbations have been derived in Refs.~\cite%
{KYY,Gao,DT11} on the flat Friedmann-Lema\^{\i}tre-Robertson-Walker (FLRW)
background in the absence/presence of matter. Imposing these conditions and
studying the background dynamics as well as the growth of density
perturbations \cite{DKT}, we can test for theoretical consistent models of
dark energy with numerous observational data.

For the unified description of modified gravitational theories, there is
another approach based on the effective field theory (EFT) of cosmological
perturbations \cite{Weinberg}-\cite{KTIJMPD}. The starting point of this
formalism is a generic action in unitary gauge that depends on the lapse
function and several geometric scalar quantities appearing in the $3+1$ ADM
decomposition on the flat FLRW background. Expanding the action up to
second-order in perturbations, the resulting linear perturbation equations
generally contain spatial derivatives higher than second order. Gleyzes 
\textit{et al.} \cite{Fedo} showed that the Horndeski theory satisfies the
conditions for absence of such higher-order derivatives by explicitly
rewriting the Lagrangian in terms of the ADM variables. The EFT of
cosmological perturbations can deal with a wide range of gravitational
theories beyond the domain of the Horndeski theory.

Models of the large-distance modification of gravity are required to recover
Newtonian gravity at short distances for the consistency with local gravity
tests in the Solar System. There are several ways to suppress the
propagation of the fifth force induced by a scalar degree of freedom $\phi $%
. One of them is the Vainshtein mechanism \cite{Vainshtein}, under which
non-linear scalar-field self interactions appearing e.g., in Galileon
gravity, lead to the decoupling of the scalar field from baryons inside the
radius much larger than the solar system \cite{GaVa}. Another is the
chameleon mechanism \cite{chameleon} applicable to $f(R)$ gravity \cite%
{fRlocal} and Brans-Dicke theory \cite{Yoko}, under which the fifth force
outside a spherically symmetric body is suppressed by the formation of a
thin shell inside the body with a large effective mass of the scalar field.

For the purpose of understanding the screening mechanism of the fifth force
in general, the equations of motion in the Horndeski theory were derived on
the spherical symmetric background \cite{Kimura,Kase,KNT}. The stability of
static and spherically symmetric vacuum solutions in the same theory was
also studied in Ref.~\cite{suyama} by considering the odd-parity mode of
perturbations (associated with tensor perturbations). The analysis of the
even-parity perturbations, which is much more involved due to a non-trivial
coupling between the scalar field and gravity, was recently performed in
Ref.~\cite{suyamaeven}. The spherically symmetric background solutions of
viable modified gravity models need to accommodate the screening mechanism
of the fifth force, while satisfying the stability conditions against
perturbations.

The EFT of modified gravity on the isotropic cosmological background allows
a possibility of dealing with the theories beyond Horndeski in a systematic
and unified way \cite{Park}-\cite{Xian}. If we try to apply a similar
formalism to the spherical symmetric background, there is another spatial
direction singled out by the ADM decomposition besides the temporal
direction. The EFT of modified gravity with the singled-out radial direction
has not been worked out yet.

There are several ways to deal with the perturbations of spherically
symmetric and static space-times. Some of the approaches monitor the metric
perturbations and they heavily rely both on the decomposition of the
perturbations into even and odd modes (under parity transformations on the
sphere)\ and on a full gauge fixing. This line of research includes the
pioneering work of Regge, Wheeler and Zerilli \cite{Regge,Zerilli}, leading
to the Regge-Wheeler equation for the odd modes and the Zerilli equation for
the even modes of general relativistic black hole perturbations. The
discussion of perturbations in the Horndeski class of theories presented in
Refs.~\cite{suyama,suyamaeven} falls into this class.

A second approach, which also heavily relies on gauge fixing, was followed
in Chandrasekhar's monumental monograph \cite{Chandrasekhar}, where general
relativistic black hole perturbations are discussed in terms of
Newman-Penrose spin coefficients and their perturbations. This approach is
based on the introduction of a Newman-Penrose tetrad and requires to solve
76 coupled differential equations in 50 independent variables.

A third approach keeps the full covariance in the discussion of the
perturbations and extensively uses gauge-invariant quantities. Such a
covariant 1+1+2 formalism was worked out for general relativistic
perturbations of the Schwarzschild space-time in Ref.~\cite{CB}, and later
it was generalized for generic locally rotationally symmetric space-times in
Ref.~\cite{Chris}. The advantages include a unified treatment of the even
and odd sectors of the perturbations in the form of wave equations holding
for both. The price to pay is a formalism involving a much larger number of
kinematical (shear, expansion, vorticity and acceleration type rather than
metric) variables, and the use of directional derivatives which do not
commute (as they are not related to coordinate derivatives), hence the need
for commutation relations, similarly as in the Newman-Penrose formalism.

We employ yet another formalism based on the $s$+1+1 decomposition, where $s$
is an arbitrary positive integer \cite{s+1+1a,s+1+1b} (developed with the
application to braneworld models in mind). This ADM inspired formalism
relies on a canonical, rather than a covariant approach. Based on a double
foliation of space-time, it is more restrictive than either of the
Newman-Penrose or the covariant 1+1+2 formalisms. The clear advantage over
them is a much lower number of variables. In comparison with the metric
perturbation formalism (for $s=2$) the number of variables is the same,
nevertheless the variables in the 2+1+1 ADM formalism carry canonical
interpretation, which is a clear virtue when it comes to the EFT approach.

In this paper, we study the EFT of modified gravity on a static and
spherically symmetric background by employing the 2+1+1 ADM formalism. In
the gravitational action, we take into account all the possible scalar
combinations constructed from geometric quantities. We show that the
Horndeski theory and its recent generalization by
Gleyzes-Langlois-Piazza-Vernizzi (GLPV) \cite{Gleyzes} can be accommodated
in our general framework, by explicitly rewriting the corresponding
Lagrangians in terms of the 2+1+1 covariant variables. The three independent
background equations of motion are derived in simple forms, which will be
useful for the study of the screening mechanism in general modified
gravitational theories.

We also obtain the second-order Lagrangian for odd-type perturbations in the
EFT framework to discuss the stability of spherically symmetric and static
vacuum solutions. We derive conditions for avoiding ghosts/Laplacian
instabilities and apply our results to both Horndeski and GLPV theories
(including a ``covariantized'' version of the original Galileon model \cite%
{Nicolis} whose coordinate derivatives are replaced by covariant
derivatives). We defer the study of the even-type perturbations to a
follow-up work due to its non-triviality and complexity.

This paper is organized as follows.

In Sec.~\ref{formalism} the basic elements of the 2+1+1 ADM decomposition
will be reviewed as a brief summary of the formalism developed in Refs.~\cite%
{s+1+1a,s+1+1b}.

In Sec.~\ref{backsec} we present a variational principle for a general
action in unitary gauge expressed in terms of scalars constructed from the
geometric quantities arising in the 2+1+1 decomposition. Varying the action
up to first order in perturbations allows us to derive three equations of
motion for the background.

In Sec.~\ref{hornsec} we express the Lagrangians of both Horndeski and GLPV
theories in terms of the variables appearing in the 2+1+1 formalism and show
that they belong to the sub-class of our general framework.

In Sec.~\ref{Gaugefix} we explore the diffeomorphism gauge freedom in
dealing with perturbations on the static and spherically symmetric
background. After choosing the unitary gauge $\delta \phi=0$, there is still
a remaining gauge degree of freedom associated with the time component of a
coordinate transformation vector $\xi^{\mu}$. We show that this residual
gauge degree of freedom does not affect the odd-type perturbations studied
in this paper.

In Sec.~\ref{perturbsec} we derive the second-order perturbed Lagrangian density 
for the odd-mode perturbations expressed in terms of a dynamical scalar variable 
and its derivatives.

In Sec.~\ref{stabsec} we discuss conditions for the absence of ghosts and
Laplacian instabilities and apply the results to both Horndeski and GLPV
theories. We also specialize our results for two covariant Galileon models.

Sec.~\ref{conclusec} is devoted to conclusions.

We use the abstract index notation throughout the paper, hence tensors
defined on the full space-time and on the 2-dimensional surface carry the
same set of Latin indices, but the latter obey certain projection
conditions. All quantities defined on the background will carry an overbar.




\section{The 2+1+1 formalism}

\label{formalism} 

We assume that the 4-dimensional space-time allows for a double foliation in
the 2+1+1 formalism, e.g., it can be foliated both by constant time
hypersurfaces $\Sigma _{t}$ and by constant spatial coordinate hypersurfaces 
$\Sigma _{r }$. The time-like unit congruence $n^{a}$ (satisfying $%
n^{a}n_{a}=-1$) is orthogonal to $\Sigma_{t} $, while the unit vector $l^{a}$
of the singled-out spatial direction (satisfying $l^{a}l_{a}=1$) is
orthogonal to $\Sigma _{r}$. For convenience we choose them mutually
orthogonal ($n^{a}l_{a}=0$). The 2-surface orthogonal to both congruences is
labeled as $\Sigma _{tr}$.

In terms of the 4-dimensional metric $g_{ab}$ and the unit vectors mentioned
above, the induced metric $h_{ab}$ on the 2-dimensional space is given by 
\begin{equation}
h_{ab}=g_{ab}+n_{a}n_{b}-l_{a}l_{b}\,,  \label{metric}
\end{equation}
which satisfies the orthogonal relations $n^{a}h_{ab}=0$ and $l^{a}h_{ab}=0$.

Time evolution proceeds along the integral lines of the congruence 
\begin{equation}
\left( \frac{\partial }{\partial t}\right) ^{a}=Nn^{a}+N^{a}\,,
\label{temporal}
\end{equation}
where $N^{a}$ and $N$ are the shift vector and lapse function related to the
foliation $\Sigma _{t}$. The singled-out spatial evolution proceeds along 
\begin{equation}
\left( \frac{\partial }{\partial r }\right) ^{a}=Ml^{a}+M^{a}\,,
\label{off-brane}
\end{equation}
where $M^{a}$ and $M$ represent the shift vector and lapse function
associated with the singled-out spatial direction. In contrast with $N$ and $%
N^{a}$, the scalar $M$ and vector $M^{a}$ represent true gravitational
degrees of freedom, contributing to the spatial 3-metric.

The shift vectors are restricted as $%
M^{a}n_{a}=M^{a}l_{a}=N^{a}n_{a}=N^{a}l_{a}=0$, so they have only two
independent components each. The gravitational sector is described by $%
\{h_{ab},M^{a},M,N^{a},N\}$, a set of one variable less than the number of
variables contained in $g_{ab}$. This is because mutually perpendicular
foliations are chosen through the condition $n^{a}l_{a}=0$.

In order to accommodate the possibility that the perturbations may affect
the perpendicularity of the foliations, we consider the 4-dimensional metric
in the system adapted to the coordinates $(t,r,x^{a})$ (here $x^{a}$ being
coordinates adapted to $\Sigma _{tr}$) in full generality \cite{s+1+1a}: 
\begin{eqnarray}
ds^{2} &=&(N_{a}N^{a}+\mathcal{N}^{2}-N^{2})dt^{2}+2N_{a}dtdx^{a}+2\left(
N_{a}M^{a}+\mathcal{N}M\right) dtdr  \notag \\
&&+h_{ab}dx^{a}dx^{b}+2M_{a}dx^{a}dr+(M_{a}M^{a}+M^{2})dr^{2}\;.  \label{ds2}
\end{eqnarray}
The requirement of a double foliation (of both vector fields $n$ and $l$
being vorticity-free), as shown in the Appendix of Ref.~\cite{s+1+1a},
imposes a proportionality of the metric functions $M$ and $\mathcal{N}$. The
easiest way to obey this constraint is to chose $\mathcal{\bar{N}}=0$,
equivalent to the perpendicularity of the foliations on the background. The
latter condition can be fulfilled even in the presence of the perturbations
by a suitable gauge fixing: 
\begin{equation}
\mathcal{N}=0\,.  \label{conditionN}
\end{equation}

The embedding of the co-dimension 2 surfaces is characterized by two types
of extrinsic curvatures, related to each of the normal vector fields $n^{a}$
and $l^{a}$: 
\begin{equation}
K_{ab}=h^{c}{}_{a}h^{d}{}_{b}\nabla _{c}n_{d}\ ,\qquad
L_{ab}=h^{c}{}_{a}h^{d}{}_{b}\nabla _{c}l_{d}\;,\   \label{extr3}
\end{equation}
where $\nabla$ denotes the $g$-metric compatible connection.

There are also two normal fundamental forms 
\begin{equation}
\mathcal{K}_{a}=h^{b}{}_{a}l^{c}\nabla _{c}n_{b}\ ,\qquad \mathcal{L}
_{a}=-h^{b}{}_{a}n^{c}\nabla _{c}l_{b}~,  \label{calK1}
\end{equation}
and two normal fundamental scalars 
\begin{equation}
\mathcal{K}=l^{a}l^{b}\nabla _{a}n_{b}\,, \qquad \mathcal{L}%
=n^{a}n^{b}\nabla _{a}l_{b}\   \label{calK2}
\end{equation}
to consider\footnote{%
The sets ($K_{ab}$, $\mathcal{K}_{a}$, $\mathcal{K}$) and ($L_{ab}$, $%
\mathcal{L}_{a}$, $\mathcal{L}$) can also be interpreted as the tensorial,
vectorial and scalar contributions in the 2+1 split of the extrinsic
curvatures of the hypersurfaces perpendicular to the congruences $n^{a}$ and 
$l^{a}$, respectively.}. To summarize, the covariant derivatives of the
normal vectors can be expressed as 
\begin{eqnarray}
\nabla _{a}n_{b} &=&K_{ab}+l_{a}\mathcal{K}_{b}+l_{b}\mathcal{K}
_{a}+l_{a}l_{b}\mathcal{K}+n_{a}\alpha _{b}\,,  \label{nablan} \\
\nabla _{a}l_{b} &=&L_{ab}+n_{a}\mathcal{L}_{b}+n_{b}\mathcal{L}
_{a}+n_{a}n_{b}\mathcal{L}+l_{a}\beta _{b}\,,  \label{nablal}
\end{eqnarray}
where $\alpha ^{a}$ and $\beta ^{a}$ are the curvatures of the congruences $%
n^{a}$ and $l^{a}$, defined by 
\begin{eqnarray}
\alpha ^{a} &=&n^{c}\nabla _{c}n^{a}=D^{a}\left( \ln N\right) -\mathcal{L}%
l^{a}\ ,  \label{acc} \\
\beta ^{a} &=&l^{c}\nabla _{c}l^{a}=-D^{a}\left( \ln M\right) +\mathcal{K}%
n^{a}\,.  \label{lambda}
\end{eqnarray}
Occasionally, both $\alpha ^{a}$ and $\beta ^{a}$ will be referred as
accelerations.

From the symmetric property of the extrinsic curvatures and the relation $%
n^{a}l_{a}=0$, it has been shown in Ref.~\cite{s+1+1a} that $\mathcal{K}_{a}=%
\mathcal{L}_{a}$ holds. The quantities $L_{ab}$ and $\mathcal{L}$ are
expressed in terms of $r$-derivatives and the covariant derivatives $D_{a}$
associated with $h_{ab} $ as \cite{s+1+1a} \label{Ls} 
\begin{eqnarray}
L_{ab} &=&\frac{1}{2M}\left( \frac{\partial h_{ab}}{\partial r}
-2D_{(a}M_{b)}\right) \,,  \label{calLabexp} \\
\mathcal{L} &=&-\frac{1}{MN}\left( \frac{\partial N}{\partial r}
-M^{a}D_{a}N\right) \,.  \label{calLexp}
\end{eqnarray}
Hence they are just convenient abbreviations for spatial derivatives.

By contrast, the quantities $K_{ab}$, $\mathcal{K}^{a}$ and $\mathcal{K}$
give the time evolution of $h_{ab}$, $M^{a}$ and $M$, respectively \cite%
{s+1+1a}: 
\begin{subequations}
\label{Ks}
\begin{eqnarray}
K_{ab} &=&\frac{1}{2N}\left( \frac{\partial h_{ab}}{\partial t}
-2D_{(a}N_{b)}\right) \ , \\
\mathcal{K}^{a} &=&\!\!\!\frac{1}{2MN}\!\left( \frac{\partial M^{a}}{
\partial t}\!-\!\frac{\partial N^{a}}{\partial r }\!+M^{b} D_{b} N^{a}
-N^{b} D_{b} M^{a} \right) \!~, \\
\mathcal{K} &=&\frac{1}{MN}\left( \frac{\partial M}{\partial t}
-N^{a}D_{a}M\right) \,,  \label{calKexp}
\end{eqnarray}
so that they are velocity-type variables.

Thus the coordinates in the velocity phase-space are 
\end{subequations}
\begin{equation}
\{h_{ab},M^{a},M;K_{ab},\mathcal{K}^{a},\mathcal{K}\}\,,  \label{vars}
\end{equation}
which is a feature any 2+1+1 covariant Lagrangian description of modified
gravity should take into account.

Note that the time and spatial derivatives along the singled-out directions
of any tensor $T_{b_{1}...b_{s}}^{a_{1}...a_{r}}$ which has vanishing
contraction with both $n^{a}$ and $l^{a}$ in all indices are defined as
projected Lie-derivatives \cite{s+1+1a,s+1+1b}: 
\begin{eqnarray}
\frac{\partial }{\partial t}T_{b_{1}...b_{s}}^{a_{1}...a_{r}} &\equiv
&h_{c_{1}}^{a_{1}}{}...h_{c_{r}}^{a_{r}}{}h_{b_{1}}^{d_{1}}{}...h_{b_{s}}^{d_{s}}~^{\left( 4\right) } 
\mathcal{L}_{\frac{\mathbf{\partial }}{{\partial t}}
}T_{d_{1}...d_{s}}^{c_{1}...c_{r}}\equiv \mathcal{L}_{\frac{\mathbf{\partial 
}}{\partial t}}T_{b_{1}...b_{s}}^{a_{1}...a_{r}}  \notag \\
&=&N\mathcal{L}_{\bm n}T_{b_{1}...b_{s}}^{a_{1}...a_{r}} +\mathcal{L}_{\bm %
N}T_{b_{1}...b_{s}}^{a_{1}...a_{r}}\ ,  \label{derivt} \\
\frac{\partial }{\partial r }T_{b_{1}...b_{s}}^{a_{1}...a_{r}} &\equiv
&h_{c_{1}}^{a_{1}}{}...h_{c_{r}}^{a_{r}}{}h_{b_{1}}^{d_{1}}{}...h_{b_{s}}^{d_{s}}~^{\left( 4\right) } 
\mathcal{L}_{\frac{\mathbf{\partial }}{\mathbf{\partial }r }
}T_{d_{1}...d_{s}}^{c_{1}...c_{r}}\equiv \mathcal{L}_{\frac{\mathbf{\partial 
}}{\mathbf{\partial }r }}T_{b_{1}...b_{s}}^{a_{1}...a_{r}}  \notag \\
&=&M\mathcal{L}_{{\bm l}}T_{b_{1}...b_{s}}^{a_{1}...a_{r}} +\mathcal{L}_{\bm %
M}T_{b_{1}...b_{s}}^{a_{1}...a_{r}}\,,  \label{derivchi}
\end{eqnarray}
where $^{\left( 4\right) }\mathcal{L}_{\bm V}$ and $\mathcal{L}_{\bm V}$
hold for the 4-dimensional and 2-dimensional Lie-derivatives along any
vector congruence ${\bm V}$. For a scalar quantity $S$, one has 
\begin{eqnarray}
\frac{\partial }{\partial t}S &=&N\left( n^{a}\nabla _{a}\right)
S+N^{a}D_{a}S\,,  \label{St} \\
\frac{\partial }{\partial r }S &=&M\left( l^{a}\nabla _{a}\right)
S+M^{a}D_{a}S\,.  \label{Schi}
\end{eqnarray}

From the above expressions, it is immediate to see that the time and spatial
derivatives along the singled-out direction of scalars which are constant on 
$\Sigma _{tr}$ (such that the last terms in Eqs.~(\ref{St}) and (\ref{Schi})
vanish) are also expressible as projected covariant derivatives, a property
we will employ in what follows. For the rest of our paper, we also denote
time derivatives by a dot and the derivatives along the singled-out spatial
direction by a prime.


\section{Equations of motion on the spherically symmetric background}

\label{backsec} 

We consider general gravitational theories with a single scalar degree of
freedom $\phi$. On the background the scalar field has only radial
dependence. As will be discussed in detail in Sec.~\ref{Gaugefix}, we choose
a radial unitary gauge $\phi=\phi(r)$. Then the kinetic term of the scalar
field can be expressed in terms of the radial lapse $M$ and the radial
derivative of the field. Hence we render the scalar field into the
gravitational sector (the radial lapse) and into the explicit radial
dependence of the action. We will therefore consider an action principle
with the Lagrangian depending on variables constructed from the metric
alone, however with explicit radial dependence allowed.

\subsection{Action principle}

We elaborate the variational principle developed for a cosmological setup 
\cite{Fedo} in a way that it applies to a spherically symmetric background.
For this purpose we employ scalar quantities related to the velocity
phase-space variables (\ref{vars}) emerging in the 2+1+1 decomposition. We
introduce the gravitational action 
\begin{equation}
S^{\mathrm{EFT}}=\int d^{4}x\sqrt{-g}\,L^{\mathrm{EFT}} \left( N,\mathcal{L}%
;M,\mathcal{K}; \mathfrak{M},\mathfrak{K}; \mathcal{R},K,\varkappa ,
L,\lambda ;r\right)\,,  \label{action}
\end{equation}
where we have denoted the gravitational Lagrangian by $L^{\mathrm{EFT}}$ and 
\begin{eqnarray}
\mathcal{R} &\equiv &{}^{(2)}{R^{a}}_{a}\,,\qquad \mathfrak{M}\equiv
M_{a}M^{a}~,\qquad \mathfrak{K}\equiv \mathcal{K}_{a} \mathcal{K}^{a}= 
\mathcal{L}_{a}\mathcal{L}^{a}~,  \notag \\
K &\equiv &{K^{a}}_{a}\,,\qquad \varkappa \equiv
K_{~b}^{a}K_{~a}^{b}\,,\qquad L\equiv {L^{a}}_{a}\,,\qquad \lambda \equiv
L_{~b}^{a}L_{~a}^{b}\,.  \label{scalardef}
\end{eqnarray}
Here ${}^{(2)}R_{ab}$ is the 2-dimensional Ricci tensor.

The action (\ref{action}) depends on the lapse and the velocity phase-space
variables (\ref{vars}) discussed in the previous section. Symmetry allows us
to use fewer variables. While the scalar sector $\{M;\mathcal{K}\}$ is fully
included, the vectorial sector $\{M^{a};\mathcal{K}^{a}\}$ appears through
the quantities $\{\mathfrak{M} ;\mathfrak{K} \}$. The tensorial sector $%
\{h_{ab};K_{ab}\}$ also appears through $\mathcal{R}$ (which in two
dimensions is the only independent component of the Riemann curvature tensor
constructed from $h_{ab}$) and the quantities $K,\varkappa $. Besides, the
scalars $\{\mathcal{L},L,\lambda \}$ formed from spatial derivatives of $N$
and $h_{ab}$ are also introduced.

In comparison to the corresponding action of the cosmological setup \cite%
{Fedo}, the action (\ref{action}) does not depend on the variable $\mathcal{Z%
}\equiv {}^{(2)}R_{ab}{}^{(2)}R^{ab}\,$, as in two dimensions the intrinsic
curvature has only one degree of freedom. In particular, the relation $^{(2)}%
{R}_{ab}=\left( \mathcal{R}/2\right) h_{ab}$ holds, so that $\mathcal{Z}=%
\mathcal{R}^{2}/2$. By contrast, the extrinsic curvatures $K_{ab} $ and $%
L_{ab}$ of the 2-dimensional surface have two independent components each
(related to the two sectional curvatures), hence we keep $\varkappa$
(denoted $\mathcal{S}$ in Ref.~\cite{Fedo}) and the new variable $\lambda$.

In summary, we have taken into account scalars equivalent to the variables
of the velocity phase-space. As the action contains a Lagrangian density $%
\sqrt{-g}\,L^{\mathrm{EFT}}$, scalars representing spatial derivatives have
been also included. Instead of the induced 2-metric $h_{ab}$, we have
included the 2-dimensional scalar curvature (in two dimensions the curvature
generated by the metric is equivalent to the scalar curvature). Finally, we
included the scalars $\varkappa ,\ \mathfrak{K}$ and $\lambda $ for later
convenience, as they also appear in the 2+1+1 version of the twice
contracted Gauss equation \cite{s+1+1a}: 
\begin{equation}
\mathcal{R}={}^{(4)}{R}+K^{2}-\varkappa -2\mathfrak{K}+2\mathcal{K}%
K-L^{2}+\lambda +2\mathcal{L}L+2\alpha ^{b}\beta _{b}+2\nabla _{a}(\alpha
^{a}-\beta ^{a}-Kn^{a}+Ll^{a})\ .  \label{curvscalar}
\end{equation}
Note that Eq.~(\ref{curvscalar}) contains a 4-dimensional covariant
derivative and is not completely written in 2-dimensional language, but it
is adequate for our work. The expression for the Ricci scalar fully
translated into 2-dimensional language can be found in Ref.~\cite{s+1+1b}%
\footnote{%
After the change in notation, $(\mathcal{R},\,{}^{\left( 4\right)
}R,\,h_{ab})\leftrightarrow (R\,,\,\tilde{R}\,,\,g_{ab})$, one can show the
equivalence between Eq.~(\ref{curvscalar}) in this paper and Eq.~(A1) in
Ref.~\cite{s+1+1b}.}.

\subsection{Background equations of motion}

In what follows, we will proceed in deriving the equations of motion by
taking variations of the action on a spherically symmetric and static
background. Under the assumption of spherical symmetry, the line element (%
\ref{ds2}) contains only two free functions of ($t$, $r$) and it simplifies
to 
\begin{equation}
d\bar{s}^{2}=-\bar{N}^{2}dt^{2}+\bar{M}^{2}dr^{2}+r^{2}d\Omega ^{2}\,,
\label{line0}
\end{equation}%
where $d\Omega ^{2}=d\theta ^{2}+(\sin ^{2}\theta )d\varphi ^{2}$ is the
surface element of the unit sphere. Since $\bar{N}^{a}=\bar{M}^{a}=0$, it
follows that $\bar{\mathfrak{M}}=\bar{\mathfrak{K}}=0$. The one-forms $n_{a}$
and $l_{a}$ are given by $n_{a}=(-N,0,0,0)$ and $l_{a}=(0,M,0,0)$,
respectively. Note that these expressions of $n_{a}$ and $l_{a}$ stay valid
to first order in the perturbations. The extrinsic curvatures obey $\bar{K}%
_{ab}=\bar{K}\bar{h}_{ab}/2$ and $\bar{L}_{ab}=\bar{L}\bar{h}_{ab}/2$, hence 
$\bar{\varkappa}=\bar{K}^{2}/2\,$ and $\bar{\lambda}=\bar{L}^{2}/2$. If the
background is further time-independent, then the relations $\bar{K}=\bar{%
\varkappa}=\mathcal{\bar{K}}=0$ hold. Other non-vanishing geometric
quantities are given by 
\begin{equation}
\bar{\mathcal{R}}=\frac{2}{r^{2}}\,,\qquad \bar{\mathcal{L}}=-\frac{\bar{N}%
^{\prime }}{\bar{N}\bar{M}}\,,\qquad \bar{L}=\frac{2}{\bar{M}r}\,,\qquad 
\bar{\lambda}=\frac{2}{\bar{M}^{2}r^{2}}\,.  \label{background}
\end{equation}

We expand the action (\ref{action}) up to second order in perturbations of
the geometric scalar quantities. In doing so, we define the variation of the
velocity phase-space variables in the action as the difference between the
background and perturbed variables. In particular, we have 
\begin{equation}
\delta \mathcal{R} \equiv \mathcal{R}-\frac{2}{r^{2}}\,,\qquad \delta 
\mathcal{L\equiv L}+\frac{\bar{N}^{\prime }}{\bar{N}\bar{M}}\,, \qquad
\delta L \equiv L-\frac{2}{\bar{M}r}\,,\qquad \delta \lambda \equiv \lambda
- \frac{2}{\bar{M}^{2}r^{2}}\ ,  \label{deltas}
\end{equation}
and 
\begin{equation}
\delta \mathfrak{K} \equiv \mathfrak{K} \,,\qquad \delta \varkappa \equiv
\varkappa\,,\qquad \delta \mathfrak{M} \equiv \mathfrak{M} \,, \qquad \delta
K \equiv K\,,\qquad \delta \mathcal{K}\equiv \mathcal{K}\,.
\end{equation}
Alternatively, from the definitions of the variables, we obtain the
following explicit expressions 
\begin{eqnarray}
\delta \lambda &=& L_{~b}^{a}L_{~a}^{b}-\bar{L}_{~b}^{a}\bar{L} _{~a}^{b}=%
\frac{2}{\bar{M}r}\delta L+\delta L_{~b}^{a}\delta L_{~a}^{b}~,  \notag \\
\delta \mathfrak{M} &=& M^{a}M_{a}-\bar{M}^{a}\bar{M}_{a}=\delta M_{a}\delta
M^{a}\,,  \notag \\
\delta \mathfrak{K} &=&\mathcal{K}^{a}\mathcal{K}_{a}-\mathcal{\bar{K}} ^{a}%
\mathcal{\bar{K}}_{a}=\delta \mathcal{K}_{a}\delta \mathcal{K}^{a}\,,  \notag
\\
\delta \varkappa &=& {K^{a}}_{b}{K^{b}}_{a} -\bar{K}_{~b}^{a} \bar{K}%
_{~a}^{b} =\delta K_{~b}^{a}\delta K_{~a}^{b}\,.
\end{eqnarray}
Hence the variables $\mathfrak{M} $, $\mathfrak{K} $ and $\varkappa$ (which
vanish on the background) are second order, while $\lambda$ (non-vanishing
on the background) is changed by the perturbations at both first and second
order. We also see that the scalar variables $\lambda$ and $L$ are not
independent at first-order accuracy.

Next we expand the Lagrangian in the action (\ref{action}) up to first order
in perturbations. In doing so, we keep in mind that $\mathfrak{M}$, $%
\mathfrak{K}$ and $\varkappa $ are second-order quantities, while at first
order $\delta \lambda $ is related to $\delta L$. This leaves us with the
following Taylor expansion: 
\begin{eqnarray}
&&L^{\mathrm{EFT}}\left( N,\mathcal{L};M,\mathcal{K};\mathfrak{M}, \mathfrak{%
K};\mathcal{R},K,\varkappa ,L,\lambda ;r\right)  \notag \\
&=&\bar{L}^{\mathrm{EFT}}+L_{N}^{\mathrm{EFT}}\delta N+ L_{\mathcal{L}}^{%
\mathrm{EFT}}\delta \mathcal{L}+L_{M}^{\mathrm{EFT}}\delta M +L_{\mathcal{K}%
}^{\mathrm{EFT}}\delta \mathcal{K} +L_{\mathcal{R}}^{\mathrm{EFT}}\delta 
\mathcal{R}+L_{K}^{\mathrm{EFT}}\delta K+\mathcal{F}\delta L\,,  \label{Lag}
\end{eqnarray}
where we introduced the notations $L_{G}^{\mathrm{EFT}}\equiv \overline{%
\partial L^{\mathrm{EFT}}/\partial G}$ for any $G=N,\mathcal{L};M,\mathcal{K}%
;\mathfrak{M}, \mathfrak{K};\mathcal{R},K,\varkappa ,L,\lambda $ (evaluated
on the background), and 
\begin{equation}
\mathcal{F}\equiv L_{L}^{\mathrm{EFT}} +\frac{2L_{\lambda}^{\mathrm{EFT}}}{ 
\bar{M}r}\,.
\end{equation}

In what follows, we explore further relations among the scalar variables. On
using Eq.~(\ref{metric}), we have that $L=h^{ab}\nabla _{a}l_{b}=\nabla
_{a}l^{a}+\mathcal{\bar{L}}+\delta \mathcal{L}$. Integrating by parts the
term $\sqrt{-g}\,\mathcal{F}\delta L$ in the action and dropping the total
covariant divergence term, finally employing Eq.~(\ref{Schi}) and the
expression (\ref{background}) of $\mathcal{\bar{L}}$, we obtain 
\begin{equation}
\int d^{4}x\sqrt{-g}\,\mathcal{F}\delta L=-\int d^{4}x\sqrt{-g} \frac{%
\mathcal{F}^{\prime }}{\bar{M}}\left( 1-\frac{\delta M}{\bar{M}}\right)
+\int d^{4}x\sqrt{-g}\,\mathcal{F}\left( -\frac{\bar{N}^{\prime }}{\bar{N} 
\bar{M}}+\delta \mathcal{L}-\frac{{2}}{r\bar{M}}\right) \,,
\end{equation}
where we have also expanded $M^{-1}$ up to first order. In the same way,
using $\delta K=K=\nabla _{a}n^{a}-\delta \mathcal{K}$, integrating by
parts, dropping the total covariant divergence term and taking into account
Eq.~(\ref{St}), it follows that 
\begin{equation}
\int d^{4}x\sqrt{-g}\,L_{K}^{\mathrm{EFT}}\delta K =-\int d^{4}x\sqrt{-g}%
\,L_{K}^{\mathrm{EFT}}\delta \mathcal{K}~.
\end{equation}
Then the Lagrangian (\ref{Lag}) is decomposed as 
\begin{equation}
L^{\mathrm{EFT}}=\bar{L}_{0}^{\mathrm{EFT}}+\delta L^{\mathrm{EFT}}\,,
\label{Lag1}
\end{equation}
where we have denoted 
\begin{equation}
\bar{L}_{0}^{\mathrm{EFT}}=\bar{L}^{\mathrm{EFT}} -\frac{\mathcal{F}^{\prime}%
}{\bar{M}}-\frac{(\bar{N}^{\prime }r +2\bar{N})\mathcal{F}}{\bar{N}\bar{M}r}%
\,,
\end{equation}
and 
\begin{equation}
\delta L^{\mathrm{EFT}}=L_{N}^{\mathrm{EFT}}\delta N+\left( L_{\mathcal{L}}^{%
\mathrm{EFT}}+\mathcal{F}\right) \delta \mathcal{L}+\left( L_{M}^{\mathrm{EFT%
}}+\frac{\mathcal{F}^{\prime }}{\bar{M}^{2}}\right) \delta M+\left( L_{%
\mathcal{K}}^{\mathrm{EFT}}-L_{K}^{\mathrm{EFT}}\right) \delta \mathcal{K}%
+L_{\mathcal{R}}^{\mathrm{EFT}}\delta \mathcal{R}\,.  \label{Lag2}
\end{equation}
It can be proven that the zeroth-order Lagrangians $\bar{L}_{0}^{\mathrm{EFT}%
}$ and $\bar{L}^{\mathrm{EFT}}$ differ only by a total covariant divergence,
which can be dropped.

The Lagrangian density is given by $\mathscr{L}=\sqrt{-g}\,L^{\mathrm{EFT}}$%
, with $\sqrt{-g}=NM\sqrt{h}$ and $\sqrt{h}=r ^{2}\sin \theta$. It can be
decomposed into a background contribution $\mathscr{\bar{L}}_{0} =\sqrt{-%
\bar{g}}\,\bar{L}^{\mathrm{EFT}}_{0}$ and a first-order contribution $\delta%
\mathscr{L}=\mathscr{L} -\mathscr{\bar{L}}_{0}$ as follows: 
\begin{equation}
\delta\mathscr{L}=\sqrt{-\bar{g}}\,\delta L^{\mathrm{EFT}} +\bar{L}^{\mathrm{%
EFT}}_{0}\,\delta \sqrt{-g}\,.  \label{Lagden1}
\end{equation}
Up to first order in perturbations the metric is given by 
\begin{eqnarray}
ds_{1}^{2} &=&-\left( \bar{N}^2+2\bar{N}\delta N\right) dt^{2} +2\bar{M}%
\delta{\mathcal{N}}dtdr+2\delta N_{a}dtdx^{a}  \notag \\
&&+\left( \bar{h}_{ab}+\delta h_{ab}\right) dx^{a}dx^{b}+2\delta
M_{a}dx^{a}dr +\left( \bar{M}^2+2\bar{M}\delta M\right) dr^{2}\,,
\label{ds1}
\end{eqnarray}
and hence 
\begin{equation}
\delta \sqrt{-g}=\frac{\sqrt{-\bar{g}}}{2}\bar{g}^{ab}\delta g_{ab} =\sqrt{-%
\bar{g}}\left( \frac{\delta N}{\bar{N}} +\frac{\delta M}{\bar{M}} +\frac{1}{2%
} \bar{h}^{ab}\delta h_{ab}\right) \,.  \label{delg}
\end{equation}

We assume the form $h_{ab}=e^{2\zeta }\bar{h}_{ab}$, where $\zeta $ is the
curvature perturbation. This is consistent with allowing only scalar
perturbations and suitably fixing the gauge, like in the cosmological case 
\cite{Fedo,LS}, see also the discussion of scalar perturbations in Sec.~\ref%
{Gaugefix}. Hence the perturbed and unperturbed metrics are related by a
conformal transformation and the respective curvature scalars can be
expressed as 
\begin{equation}
\mathcal{R}=e^{-2\zeta }\left( \mathcal{\bar{R}}-2\bar{h}^{ab} \bar{D}_{a} 
\bar{D}_{b}\zeta \right),  \label{Rperturb}
\end{equation}
which to linear order gives 
\begin{equation}
\delta \mathcal{R}=-2\zeta \mathcal{\bar{R}}-2\bar{h}^{ab} \bar{D}_{a}\bar{D}%
_{b}\zeta \,.  \label{delR}
\end{equation}
In the generalized Stokes theorem, the integral of a differential form $%
\omega $ over the boundary of an oriented manifold $\mathcal{S}$ is
equivalent to the integral of the exterior derivative of $\omega $ over the
manifold $\mathcal{S}$, i.e. $\int_{\mathcal{S}}d\omega =\int_{\partial 
\mathcal{S}}\omega $. Since there is no boundary of a boundary, the rhs of
the generalized Stokes theorem vanishes when $\mathcal{S}$ is some closed
surface, e.g., the 2-sphere as in our case. Using this and integrating the
second term on the rhs of Eq.~(\ref{delR}), we obtain 
\begin{equation}
\int d^{4}x\sqrt{-g}\,L_{\mathcal{R}}^{\mathrm{EFT}}\left( -2\bar{h}^{ab}%
\bar{D}_{a}\bar{D}_{b}\zeta \right) =-2\int dtdr\bar{N}\bar{M}r^{2}L_{%
\mathcal{R}}^{\mathrm{EFT}}\int d\theta \,d\varphi D_{a}\left( \sqrt{\bar{h}}%
\bar{h}^{ab}\bar{D}_{b}\zeta \right) =0\,.
\end{equation}%
Hence the variations in the scalar curvature and conformal factor are
related by the simple expression 
\begin{equation}
\delta \mathcal{R}=-\frac{4\zeta }{r^{2}}\,.  \label{delRsphsymm}
\end{equation}%
Remarkably, the same expression emerges for restricting to spherically
symmetric perturbations. Non-spherically symmetric modes in the
perturbations do not contribute to the background equations of motion.

Similarly, to linear order in perturbations, we obtain 
\begin{equation}
\frac{1}{2}\bar{h}^{ab}\delta h_{ab}=2\zeta \,,  \label{delh}
\end{equation}
which, when employing Eq.~(\ref{delRsphsymm}) and the first equation (\ref%
{background}), becomes 
\begin{equation}
\frac{1}{2}\bar{h}^{ab}\delta h_{ab}=-\frac{\delta \mathcal{R}} {\bar{%
\mathcal{R}}}\,.  \label{delh2}
\end{equation}
With this, we have completed the program of rewriting the linear-order
variation exclusively into terms containing the variation of the scalar
variables in the action.

In what follows we further reduce this set at linear order. Substitution of
Eqs.~(\ref{delg}) and (\ref{delh2}) into the first-order Lagrangian density (%
\ref{Lagden1}) leads to 
\begin{eqnarray}
\delta\mathscr{L} &=& \sqrt{-\bar{g}}\left[ L^{\mathrm{EFT}}_{N}\delta
N+\left( L^{\mathrm{EFT}}_{\mathcal{L}}+\mathcal{F}\right) \delta \mathcal{L}
+\left( L^{\mathrm{EFT}}_{M} +\frac{\mathcal{F}^{\prime }}{\bar{M}^{2}}
\right) \delta M +\left( L^{\mathrm{EFT}}_{\mathcal{K}}-L^{\mathrm{EFT}
}_{K}\right) \delta \mathcal{K} +L^{\mathrm{EFT}}_{\mathcal{R}}\delta 
\mathcal{R} \right]\,  \notag \\
&&+\bar{L}^{\mathrm{EFT}}_{0}\sqrt{-\bar{g}}\left( \frac{\delta N}{\bar{N}} +%
\frac{\delta M}{\bar{M}} -\frac{\delta \mathcal{R}}{\bar{\mathcal{R}}}
\right) \,.  \label{preLagden}
\end{eqnarray}
By using Eqs.~(\ref{calLexp}) and (\ref{calKexp}), it follows that 
\begin{eqnarray}
\delta \mathcal{L}&=&\frac{\bar{N}^{\prime }}{\bar{N}\bar{M}} \left(-\frac{%
\delta N^{\prime }}{\bar{N}^{\prime }} +\frac{\delta N}{\bar{N}} +\frac{%
\delta M}{\bar{M}}\right)\,,  \label{delcalL} \\
\delta \mathcal{K}&=&\frac{\dot{\delta M}}{\bar{N}\bar{M}}\,.
\label{delcalK}
\end{eqnarray}
Plugging these expressions into Eq.~(\ref{preLagden}) and integrating by
parts, we obtain 
\begin{eqnarray}
\delta\mathscr{L} &=& \sqrt{-\bar{g}} \Bigg[ \left\{ L^{\mathrm{EFT}}_{N}+ 
\frac{\left( L^{\mathrm{EFT}}_{\mathcal{L}}+\mathcal{F}\right) ^{\prime }} {%
\bar{M}\bar{N}}+\frac{( \bar{N}^{\prime }r+2\bar{N})\left( L^{\mathrm{EFT}}_{%
\mathcal{L}}+\mathcal{F}\right)} {\bar{N}^{2}\bar{M}r}\right\} \delta
N+\left\{ L^{\mathrm{EFT}}_{M}+\frac{\mathcal{F} ^{\prime}} {\bar{M}^{2}}+%
\frac{\bar{N}^{\prime }\left( L^{\mathrm{EFT}}_{\mathcal{L}}+ \mathcal{F}
\right)}{\bar{N}\bar{M}^{2}}\right\} \delta M  \notag \\
&&~~~~~~~~ +L^{\mathrm{EFT}}_{\mathcal{R}}\delta \mathcal{R}\Bigg] +\bar{L}^{%
\mathrm{EFT}}_{0} \sqrt{-\bar{g}}\left( \frac{\delta N}{\bar{N}} +\frac{
\delta M}{\bar{M}}-\frac{\delta \mathcal{R}}{\bar{\mathcal{R}}}\right)\,.
\end{eqnarray}
Variation of the three scalars $\delta N$, $\delta M$, and $\delta \mathcal{R%
}$ leads, respectively, to 
\begin{eqnarray}
& & \bar{L}^{\mathrm{EFT}}+\bar{N}L^{\mathrm{EFT}}_{N} +\frac{(\bar{N}%
^{\prime }r +2\bar{N})L^{\mathrm{EFT}}_{\mathcal{L}}} {\bar{N}\bar{M}r}+%
\frac{{L^{\mathrm{EFT}}_{\mathcal{L}}} ^{\prime }} {\bar{M}}=0\,,
\label{bg00} \\
& & \bar{L}^{\mathrm{EFT}}+\bar{M}L^{\mathrm{EFT}}_{M} -\frac{2\mathcal{F}}{%
\bar{M}r} +\frac{\bar{N}^{\prime } L^{\mathrm{EFT}}_{\mathcal{L}}}{\bar{M} 
\bar{N}}=0\,,  \label{bg11} \\
& & \bar{L}^{\mathrm{EFT}}-\frac{\mathcal{F}^{\prime }}{\bar{M}} -\frac{(%
\bar{N}^{\prime }r+2\bar{N})\mathcal{F}}{\bar{N}\bar{M}r} -\frac{2L^{\mathrm{%
EFT}}_{\mathcal{R}}}{r^{2}}=0\,,  \label{bg22}
\end{eqnarray}
which are the equations of motion on the spherically symmetric and static
background. For a given Lagrangian, they can be used for discussing the
screening mechanism of the fifth force mediated by the scalar degree of
freedom. In Appendix \ref{sec:eomhorn}, we show that, in the Horndeski
theory, the background equations of motion following from Eqs.~(\ref{bg00})-(%
\ref{bg22}) coincide with those derived in Refs.~\cite{Kase,suyama} by the
direct variation of the Horndeski action. In doing so, we need to express
the Horndeski action in terms of the variables used in the 2+1+1
decomposition. In the next section we shall address this issue in both
Horndeski and GLPV theories.



\section{2+1+1 decomposition of Horndeski and GLPV theories}

\label{hornsec} 

In what follows, we prove that, assuming a spherically symmetric and static
background, both the Horndeski theory \cite{Horndeski} and its recent GLPV 
\cite{Gleyzes} generalization are accommodated in the framework of the EFT
of modified gravity.

In unitary gauge, the unit normal vector orthogonal to the constant $\phi$
hypersurfaces (which coincide with the constant $r$ hypersurfaces) can be
expressed as 
\begin{equation}
l_{a}=\gamma \nabla _{a}\phi \,,\qquad \gamma =\frac{1}{\sqrt{X}}\,.
\end{equation}
By virtue of Eq.~(\ref{nablal}), the covariant derivative of $\nabla_{a}\phi
=\gamma ^{-1}l_{a}$ reads 
\begin{equation}
\nabla _{a}\nabla _{b}\phi =\gamma ^{-1}\left( L_{ab}+n_{a}\mathcal{L}
_{b}+n_{b}\mathcal{L}_{a}+n_{a}n_{b}\mathcal{L}+l_{a}\beta _{b}+l_{b}\beta
_{a}\right) +\frac{\gamma ^{2}}{2}\nabla ^{c}\phi \nabla _{c}Xl_{a}l_{b}\,.
\label{ddphi}
\end{equation}
Finally, the term $\square \phi =g^{ab}\nabla _{a}\nabla _{b}\phi$ becomes 
\begin{equation}
\square \phi =\gamma ^{-1}(L-\mathcal{L}) +\frac{\nabla ^{c}\phi \nabla
_{c}X }{2X}\,.  \label{squ}
\end{equation}
With the help of these formulas, we will rewrite both the Horndeski and GLPV
Lagrangians in terms of the 2+1+1 variables of the action (\ref{action}).

\subsection{The Horndeski class of theories}

\label{horn2+1+1}

The most general scalar-tensor theories with second-order equations of
motion~\cite{Horndeski} can be given as a series of the Lagrangians \cite%
{Deffayet11} 
\begin{equation}
L^{\mathrm{H}}=\sum_{i=2}^{5}L_{i}^{\mathrm{H}}\,,
\end{equation}
where 
\begin{eqnarray}
&&L_{2}^{\mathrm{H}}=G_{2}(\phi ,X)\,,  \label{G2} \\
&&L_{3}^{\mathrm{H}}=G_{3}(\phi ,X)\square \phi \,,  \label{G3} \\
&&L_{4}^{\mathrm{H}}=G_{4}(\phi ,X)R-2G_{4X}(\phi ,X)\left[ (\square \phi
)^{2}-\nabla ^{a}\nabla ^{b}\phi \nabla _{a}\nabla _{b}\phi \right] \,,
\label{G4} \\
&&L_{5}^{\mathrm{H}}=G_{5}(\phi ,X)G_{ab}\nabla ^{a}\nabla ^{b}\phi +\frac{1%
}{3}G_{5X}(\phi ,X)\left[ (\square \phi )^{3}-3(\square \phi )\nabla
^{a}\nabla ^{b}\phi \nabla _{a}\nabla _{b}\phi +2\nabla _{a}\nabla _{b}\phi
\nabla ^{c}\nabla ^{b}\phi \nabla _{c}\nabla ^{a}\phi \right] \,.  \label{G5}
\end{eqnarray}
Here $G_{2,3,4,5}$ are functions of a scalar field $\phi $ and of its
kinetic term $X\equiv \nabla ^{a}\phi \nabla _{a}\phi $.

The analysis of the background gravitational dynamics in the Horndeski
theory have been presented in Refs.~\cite{Kimura,KNT,Kase} on the
spherically symmetric space-time and specialized for the weak gravity
regime, allowing for confrontation with solar-system tests. In the presence
of non-linear scalar-field self interactions, the Vainshtein mechanism can
be efficient enough to suppress the propagation of the fifth force inside
the solar system, provided that the non-minimal derivative coupling to the
Einstein tensor is suppressed \cite{Kimura,KNT,Kase}. At a technical level,
this translates into constraining the magnitude of the function $G_{5}$ in
the $L_{5}^{\mathrm{H}}$ contribution of the Horndeski Lagrangian to be
subdominant as compared to the $L_{4}^{\mathrm{H}}$ contribution. For the
consistency with solar-system tests, we will consider the subclass of the
Horndeski theory with $L_{5}^{\mathrm{H}}=0$ in the following.

The Lagrangian $L_{2}^{\mathrm{H}}$ depends on the lapse $M$ according to 
\begin{equation}
L_{2}^{\mathrm{H}}=G_{2}(\phi ,X(M))\,,\qquad X(M)=\frac{\phi ^{\prime 2}}{%
M^{2}}\,.  \label{non-minimal}
\end{equation}%
As for the Lagrangian $L_{3}^{\mathrm{H}}=G_{3}\square \phi $, we introduce
an auxiliary function $F_{3}(\phi ,X)$ \cite{Fedo} such that 
\begin{equation}
G_{3}\equiv F_{3}+2XF_{3X}\,.
\end{equation}%
Integrating the term $F_{3}\square \phi $ by parts and using Eq.~(\ref{squ})
for the term $2XF_{3X}\square \phi $, the Lagrangian $L_{3}^{\mathrm{H}}$
reduces to 
\begin{equation}
L_{3}^{\mathrm{H}}=2X^{3/2}F_{3X}(L-\mathcal{L})-F_{3\phi }X\,.
\label{L3adm}
\end{equation}%
By using Eqs.~(\ref{curvscalar}), (\ref{ddphi}) and (\ref{squ}), the
Lagrangian $L_{4}^{\mathrm{H}}$ can be expressed as 
\begin{equation}
L_{4}^{\mathrm{H}}=G_{4}\left( \mathcal{R}-K^{2}+\varkappa \right) +\left(
G_{4}-2XG_{4X}\right) \left[ L^{2}-\lambda -2L\mathcal{L}+2\mathfrak{K}-2K%
\mathcal{K}+2D^{a}\left( \ln {N}\right) D_{a}\left( \ln {M}\right) \right] +2%
\sqrt{X}G_{4\phi }(L-\mathcal{L})\,.  \label{L4adm}
\end{equation}%
Thus we have shown that the Horndeski Lagrangians $L_{2,3,4}^{\mathrm{H}}$
are fully expressed in terms of 2+1+1 covariant quantities introduced in the
action~(\ref{action}).

\subsection{GLPV theories}

We proceed to apply our formalism to the GLPV Lagrangian 
\begin{equation}
L^{\mathrm{GLPV}}=\sum_{i=2}^{5}L_{i}^{\mathrm{GLPV}}\,,
\end{equation}
where the series of Lagrangians $L_{2-5}^{\mathrm{GLPV}}$ are given by \cite%
{Gleyzes} 
\begin{eqnarray}
L_{2}^{\mathrm{GLPV}} &=&A_{2}(\phi ,X)\,,  \label{exhornLa2} \\
L_{3}^{\mathrm{GLPV}} &=&\left[ C_{3}(\phi ,X)+2XC_{3X}(\phi ,X)\right]
\square \phi +XC_{3\phi }(\phi ,X)\,,  \label{exhornLa3} \\
L_{4}^{\mathrm{GLPV}} &=&B_{4}(\phi ,X)R-\frac{B_{4}(\phi ,X) +A_{4}(\phi ,X)%
}{X}\left[ (\square \phi )^{2}-\nabla ^{a} \nabla ^{b}\phi \nabla
_{a}\nabla_{b}\phi \right]  \notag \\
&&+\frac{2\left[ B_{4}(\phi ,X)+A_{4}(\phi ,X)-2XB_{4X}(\phi ,X)\right] }{%
X^{2}}\left( \nabla ^{a}\phi \nabla ^{b}\phi \nabla _{a}\nabla _{b}\phi
\square \phi -\nabla ^{a}\phi \nabla _{a}\nabla _{b}\phi \nabla _{c}\phi
\nabla ^{b}\nabla ^{c}\phi \right)  \notag \\
&&+\left[ C_{4}(\phi ,X)+2XC_{4X}(\phi ,X)\right] \square \phi +XC_{4\phi
}(\phi ,X)\,,  \label{exhornLa4} \\
L_{5}^{\mathrm{GLPV}} &=&G_{5}(\phi ,X)G_{ab}\nabla ^{a}\nabla ^{b}\phi
-|X|^{3/2}A_{5}(\phi ,X)\left[ (\square \phi )^{3}-3(\square \phi )\nabla
^{a}\nabla ^{b}\phi \nabla _{a}\nabla _{b}\phi +2\nabla _{a}\nabla _{b}\phi
\nabla ^{c}\nabla ^{b}\phi \nabla _{c}\nabla ^{a}\phi \right]  \notag \\
&&+\frac{XB_{5X}(\phi ,X)+3A_{5}(\phi ,X)}{|X|^{5/2}}\Big[\left( \square
\phi \right) ^{2}\nabla _{a}\phi \nabla ^{a}\nabla ^{b}\phi \nabla _{b}\phi
-2\square \phi \nabla _{a}\phi \nabla ^{a}\nabla ^{b}\phi \nabla _{b}\nabla
_{c}\phi \nabla ^{c}\phi  \notag \\
&&-\nabla _{a}\nabla _{b}\phi \nabla ^{a}\nabla ^{b}\phi \nabla _{c}\phi
\nabla ^{c}\nabla ^{d}\phi \nabla _{d}\phi +2\nabla _{a}\phi \nabla
^{a}\nabla ^{b}\phi \nabla _{b}\nabla _{c}\phi \nabla ^{c}\nabla ^{d}\phi
\nabla _{d}\phi \Big]  \notag \\
&&+C_{5}(\phi ,X)R-2C_{5X}(\phi ,X)\left[ (\square \phi )^{2}-\nabla
^{a}\nabla ^{b}\phi \nabla _{a}\nabla _{b}\phi \right] \,,  \label{exhorn}
\end{eqnarray}
where 
\begin{equation}
C_{3}=\int dX\frac{A_{3}}{2|X|^{3/2}}\,,\quad C_{4}=-\int dX\frac{B_{4\phi}}{%
|X|}\,,\quad C_{5}=\frac{XG_{5\phi }-|X|^{1/2}B_{5\phi }}{2}\,,\quad
G_{5}=-\int dX\frac{B_{5X}}{|X|^{1/2}}\,,
\end{equation}
with $A_{2,3,4,5}$ and $B_{4,5}$ arbitrary functions of a scalar field $\phi$
and its kinetic term $X$. The Lagrangians (\ref{exhornLa2})-(\ref{exhorn})
arise as an extension of the Horndeski theory by generalizing the Horndeski
Lagrangians written in terms of the ADM variables in the isotropic
cosmological setup \cite{Gleyzes}.

The Horndeski theory corresponds to 
\begin{eqnarray}
A_{4} &=&-B_{4}+2XB_{4X}\,,  \label{H1} \\
A_{5} &=&-\frac{XB_{5X}}{3}\,,  \label{H2}
\end{eqnarray}
under which the terms on the second line of Eq.~(\ref{exhornLa4}) and those
in the second and third lines of Eq.~(\ref{exhorn}) vanish. Then, the
Horndeski Lagrangians (\ref{G2})-(\ref{G5}) can be recovered by moving some
of the terms (such as $XC_{3\phi}(\phi,X)$) in the Lagrangian $L_{i}^{%
\mathrm{\ GLPV}}$ ($i=3,4,5$) to the previous Lagrangian $L_{i-1}^{\mathrm{%
GLPV}}$.

In comparison to the Horndeski Lagrangians characterized by the functions $%
G_{2,3,4,5}$, the theories (\ref{exhorn}) have two additional functions
included in $A_{2,3,4,5}$ and $B_{4,5}$. Apparently, the equation of motion
for the scalar field allows for derivatives higher than second order. In the
presence of higher-order derivatives\footnote{%
Although such a higher-order dynamics is non-standard in physics, it has not
been unaccounted either. An example for such a dynamics is provided by the
(spin-orbit contribution to the) Lagrangian of spinning binary black holes.
In this case the Lagrangian depends on the relative acceleration of the
black holes, which leads to a third-order Euler-Lagrange equation \cite{KWW}.%
}, the theory can be plagued by Ostrogradski instabilities associated with
the propagation of the extra degrees of freedom \cite{Ostro}. In the GLPV
theory, however, a careful counting of the degrees of freedom in the
Hamiltonian formulation on the isotropic cosmological background\footnote{%
In Ref.~\cite{Gleyzes} this has been performed after the scalar degree of
freedom is transferred into the lapse and coordinate associated with the
constant $\phi $ hypersurfaces. For the spatial hypersurfaces considered
there, the usual lapse $N$ and the time $t$ were employed. On the
spherically symmetric background the constant $\phi $-surfaces have
spherical topology, so in this case the scalar degree of freedom is
transferred into $M$ and $r$.} indicates that no additional degrees of
freedom would arise.

As in the discussion of the Horndeski theory, we will also drop the
contribution of $L_{5}^{\mathrm{GLPV}}$. The Lagrangians $L_{2,3,4}^{\mathrm{%
GLPV}}$ can be expressed as 
\begin{eqnarray}
L_{2}^{\mathrm{GLPV}} &=&A_{2}\,,  \label{exhornL2} \\
L_{3}^{\mathrm{GLPV}} &=&A_{3}\left( L-\mathcal{L}\right) \,,
\label{exhornL3} \\
L_{4}^{\mathrm{GLPV}} &=&B_{4}\left( \mathcal{R}-K^{2}+\varkappa \right)
-2\left( B_{4}-2XB_{4X}\right) \left[ K\mathcal{K}-D^{a}\left( \ln {N}
\right) D_{a}\left( \ln {M}\right) \right] -A_{4}\left( L^{2}-\lambda -L 
\mathcal{L}+2\mathfrak{K}\right) \,,  \label{exhornL4}
\end{eqnarray}
fully rewritten in terms of the 2+1+1 covariant variables of the action (\ref%
{action}). Hence $L_{2,3,4}^{\mathrm{GLPV}}$ also belong to the class of the
EFT of modified gravity. This illustrates that the latter accommodates
theories beyond Horndeski.

In Appendix~\ref{sec:eomhorn} we show the background equations of motion, as
derived from Eqs.~(\ref{bg00})-(\ref{bg22}) for the GLPV Lagrangians (\ref%
{exhornL2})-(\ref{exhornL4}). Under the conditions (\ref{H1}) and (\ref{H2}%
), the equations of motion coincide\footnote{%
In order to manifestly see this, one has to redefine the functions $A_{i}$
and $B_{i}$. These redefinitions will be discussed in Appendix A in the case
when $L_{5}^{\mathrm{GLPV}}$ is dropped.} with those derived in Refs.~\cite%
{Kase,suyama} in the Horndeski theory. In general, however, they differ from
each other.

Thus we have shown that there are theories which at the level of the
background are second order and more generic than the Horndeski theory. This
seems to contradict the generic claim that the Horndeski theory represents
the most generic second-order scalar-tensor dynamics. We have to keep in
mind however that we are considering a spherically symmetric and static
background. These additional symmetries may render some of the requirements
imposed in order to achieve second-order dynamics unnecessarily restrictive.

Further, we comment that, under spherical symmetry and staticity imposed in
the generic EFT of modified gravity, the tensorial sector is always governed
by second-order dynamics. As we consider a static background, the equations
of motion (\ref{bg00})-(\ref{bg22}) represent constraints, containing no
time derivatives. Due to the additional spherical symmetry, higher-order
derivative terms could emerge only as radial derivatives. This could happen,
if the Lagrangian $L^{\mathrm{EFT}}$ involves second radial derivatives.
Nevertheless, this is forbidden by the very nature of the action. Indeed,
the Lagrangian only depends on scalars constructed algebraically from the
variables of the 2+1+1 formalism involving the induced metric, extrinsic
curvatures, normal fundamental vectors and forms. The latter are related to
first temporal and radial derivatives, as Eqs.~(\ref{calLabexp})-(\ref%
{calKexp}) explicitly show. No second-order derivatives of the metric are
included in these variables. Hence the background equations of motion
(increasing the differential order of the Lagrangian at most by one) are
free from third or higher order radial derivatives of the chosen variables
of the action.

Nevertheless, at the level of perturbations, to be discussed in the rest of
the paper, their second-order evolution cannot be guaranteed a priori.


\section{Gauge transformations and fixing}
\label{Gaugefix} 

In this section we discuss the simplifications achieved by suitably
employing the available gauge degrees of freedom (diffeomorphism
invariance). In doing so, we will adapt the radial coordinate $r$ to the
hypersurfaces of constant scalar field even in the perturbed case by
requiring 
\begin{equation}
\delta \phi =0\,.
\end{equation}
Next, we will simplify the perturbations of the induced 2-metric to a mere
conformal rescaling. Finally we will adopt a gauge which maintains the
geometrical interpretation of the variables as arising in the 2+1+1
canonical formalism (e.g., assure $\mathcal{N}=0$ even in the presence of
perturbations).

In a manner analogous to the Helmholtz theorem, any vector $%
V_{a}=V_{a}\left(t,{r},\theta ,\varphi \right)$ on a sphere can be
decomposed by using scalar potentials as follows: 
\begin{equation}
V_{a}=\bar{D}_{a}V_{\mathrm{rot}}+{E^{b}}_{a}\bar{D}_{b}V_{\mathrm{div}}\,,
\label{vectordec}
\end{equation}
where $V_{\mathrm{rot}}=V_{\mathrm{rot}}\left( t,r,\theta ,\varphi \right)$
and $V_{\mathrm{div}}=V_{\mathrm{div}}\left( t,r,\theta ,\varphi \right)$
are arbitrary scalars generating a rotation-free part and a divergence-free
part, respectively. Here $E_{ab}=\sqrt{\bar{h}}\,\varepsilon _{ab}$ and $%
\varepsilon _{ab}$ stands for the antisymmetric tensor density, defined as $%
\varepsilon _{\theta \varphi }=1$ \cite{suyama}. Similarly, any rank-2
symmetric tensor $T_{ab}=T_{ab}\left( t,r,\theta ,\varphi \right) $ on a
sphere can be decomposed in terms of a scalar and a vector potential, e.g., $%
T_{\mathrm{scalar}}$ and $T_{a}$, as $T_{ab}= \bar{h}_{ab}T_{\mathrm{scalar}%
} +\left( \bar{D}_{a}T_{b}+\bar{D}_{b}T_{a}\right) /2$. Applying the
decomposition (\ref{vectordec}) to $T_{a}$, the tensor $T_{ab}$ is uniquely
expressed in terms of the scalar functions $T_{\mathrm{scalar}}$, $T_{%
\mathrm{rot}}$ and $T_{\mathrm{div}}$, as 
\begin{equation}
T_{ab}=\bar{h}_{ab}T_{\mathrm{scalar}}+\bar{D}_{a}\bar{D}_{b}T_{\mathrm{rot}
}+\frac{1}{2}\left( {E^{c}}_{a}\bar{D}_{c}\bar{D}_{b}+{E^{c}}_{b}\bar{D}_{c} 
\bar{D}_{a}\right) T_{\mathrm{div}}\,.  \label{tensordec}
\end{equation}

We apply these decompositions to the metric perturbation (\ref{ds1}), such
that the perturbed quantities can be expressed as 
\begin{subequations}
\label{perturbdec}
\begin{eqnarray}
\delta N_{a} &=&\bar{D}_{a}P+{E^{b}}_{a}\bar{D}_{b}Q\,,  \label{Nadec} \\
\delta M_{a} &=&\bar{D}_{a}V+{E^{b}}_{a}\bar{D}_{b}W\,,  \label{Madec} \\
\delta h_{ab} &=&\bar{h}_{ab}A+\bar{D}_{a}\bar{D}_{b}B+\frac{1}{2}\left(
{E^{c}}_{a}\bar{D}_{c}\bar{D}_{b}+{E^{c}}_{b}\bar{D}_{c} \bar{D}_{a}\right)
C\,.  \label{hdec}
\end{eqnarray}
Here the perturbations $Q$, $W$ and $C$ correspond to either divergence-free
terms or to derivatives of such terms (these terms have non-vanishing
curls), whereas $P$, $V$, $A$, and $B$ represent either rotation-free terms
or derivatives of such terms. As first shown in Ref.~\cite{Regge}, after
expanding in terms of spherical harmonics, the elements of the first set
become odd modes under the parity transformation on the sphere. The
quantities of the second set, together with $\delta N$, 
$\delta \mathcal{N}$, 
and $\delta M$ of Eq.~(\ref{ds1}), behave as even modes.

In what follows we concentrate on the evolution of these $10$ variables,
conveniently characterizing the perturbations from the parity point of view.
At first, we remark that some of them could be eliminated by making use of
the allowed diffeomorphism freedom. In doing so, we consider an
infinitesimal coordinate transformation $\tilde{x}^{a}=x^{a}+\xi^{a}$. For
the infinitesimal displacement $\xi ^{a}$ we write the time and radial
component as $\xi ^{t}$ and $\xi ^{r}$ respectively, while the infinitesimal
displacement along the sphere is decomposed as 
\end{subequations}
\begin{equation}
\xi^{a}=\bar{D}^{a}\xi +E^{ba}\bar{D}_{b}\eta\,, \qquad \left(
a=\theta,\varphi \right)\,.
\end{equation}
Then, the perturbed metric in the new coordinate system becomes $\widetilde{%
\delta g_{ab}}= \delta g_{ab}+\nabla _{a}\xi _{b}+\nabla _{b}\xi _{a}$.

The perturbations transform as 
\begin{subequations}
\label{gaugetrans}
\begin{eqnarray}
\widetilde{\delta N} &=&\delta N-\bar{N}\dot{\xi ^{t}}-\bar{N^{\prime }}\xi
^{r}\,, \\
\widetilde{\delta \mathcal{N}} &=&\delta \mathcal{N}-\frac{\bar{N}^{2}}{2%
\bar{M}}{\xi ^{t}}^{\prime }+\frac{\bar{M}}{2}\dot{\xi ^{r}}\,, \\
\widetilde{\delta M} &=&\delta M+\bar{M}^{\prime }{\xi ^{r}}+\bar{M}{\xi ^{r}%
}^{\prime }\,, \\
\widetilde{P} &=&P-\bar{N}^{2}{\xi ^{t}}+\dot{\xi}\,, \\
\widetilde{Q} &=&Q+\dot{\eta}\,, \\
\widetilde{V} &=&V+{\bar{M}^{2}}{\xi ^{r}}+{\xi }^{\prime }-\frac{2}{{r}}\xi
\,, \\
\widetilde{W} &=&W+{\eta }^{\prime }-\frac{2}{r}\eta \,, \\
\widetilde{A} &=&A+\frac{2}{r}\xi ^{r}\,, \\
\widetilde{B} &=&B+2\xi \,, \\
\widetilde{C} &=&C+2\eta \,.
\end{eqnarray}
Additionally, the linear perturbation $\delta \phi $ of a scalar field $\phi
(t,{r},\theta ,\varphi )=\bar{\phi}(r)+\delta \phi (t,{r},\theta ,\varphi )$
transforms under an infinitesimal coordinate transformation as 
\end{subequations}
\begin{equation}
\widetilde{\delta \phi }=\delta \phi -\bar{\phi}^{\prime }\xi ^{r}\,.
\end{equation}

In the isotropic cosmological setting, the key ingredient in deriving the
EFT of modified gravity is the $3+1$ decomposition with the time slicing
determined by hypersurfaces of the uniform scalar field \cite{Fedo}. In an
analogous way, we consider here the hypersurfaces of constant $\phi$ as
defining the radial slicing with ${r=}$ const, in a choice which simplifies
the EFT of modified gravity on the spherically symmetric background.
Therefore, we first fix the gauge $\xi ^{r}$ to obtain $\widetilde{\delta
\phi}=0$. Due to this gauge choice, the action (\ref{action}) does not
explicitly include the scalar field as a variable.

Next, we fix the two gauge degrees of freedom $\xi $ and $\eta $ such that
the anisotropic contributions to $\delta h_{ab}$ disappear, i.e., $%
\widetilde{B}=\widetilde{C}=0$. By doing so, the perturbed and unperturbed
induced metrics are simply related by a conformal transformation as ${h}%
_{ab}=(1+\widetilde{A})\bar{h}_{ab}$. After redefining $\widetilde{A}%
=e^{2\zeta }-1$, the perturbed induced metric coincides with the one
employed in Sec.~\ref{backsec}. Finally, we also need to fix the gauge $%
\xi^{t}$ to achieve $\widetilde{\delta \mathcal{N}}=0$ [see Eq.~(\ref%
{conditionN})]\footnote{%
Even if we would not choose $\delta \mathcal{N}=0$, preserving at the level
of perturbations the more general linear relation between $\mathcal{N}$ and $%
M$, Eq.~(C2) of the Appendix C of Ref.~\cite{s+1+1a} would consume this
gauge degree of freedom.}.

In summary, the gauge fixing is given by 
\begin{equation}
\xi^{t}=\int dr\frac{2\bar{M}}{\bar{N}^{2}}\left( \delta \mathcal{N} 
+\frac{\bar{M}}{2}\dot{\xi}^{r}\right) +F(t,\theta ,\varphi )\,,\qquad \xi ^{r}= 
\frac{\delta \phi }{\bar{\phi}^{\prime }}\,,\qquad \xi =-\frac{B}{2}
\,,\qquad \eta =-\frac{C}{2}\,,  \label{gauge}
\end{equation}
where $F(t,\theta ,\varphi )$ is an integration function, yet to be fixed
\footnote{
In the particular case where $P$ exhibits the radial dependence 
$P(t,{r},\theta ,\varphi )=\bar{N}\left( {r}\right)^{2}F(t,\theta ,\varphi )$, the
remaining gauge transformation $\tilde{t}=t+F(t,\theta ,\varphi )$ could be
employed to eliminate $\widetilde{P}$. In general, however, this is not
possible, so another fixing of the function $F$ would be necessary in order
to avoid the appearance of any non-physical gauge mode, similar to the one
of the synchronous gauge in cosmology.}.

With the new notation for the conformal factor in the transformation of the
induced metric 
\begin{equation}
\delta h_{ab}=\left( e^{2\zeta }-1\right) \bar{h}_{ab}\,,
\label{perturbtens}
\end{equation}
the line element up to first-order accuracy can be written as 
\begin{equation}
ds_{1}^{2}=-\left( \bar{N}^{2}+2\bar{N}\delta N\right) dt^{2}+2\delta
N_{a}dtdx^{a}+2\delta M_{a}dx^{a}d{r} +\left( \bar{M}^{2}+2\bar{M}\delta
M\right) d{r}^{2}+e^{2\zeta }\bar{h} _{ab}dx^{a}dx^{b}\,,  \label{metric1}
\end{equation}
where $\delta N_{a}$ and $\delta M_{a}$\ are given in terms of
parity-related scalars through Eqs.~(\ref{Nadec}) and (\ref{Madec}). In the
above expression we have omitted the tildes for notational simplicity, and
we will do so hereafter.

We now discuss how the gauge fixing affects the even and odd modes. First,
we stress that the residual gauge freedom in $F$ does not affect the
odd-parity perturbations as it does not appear in the transformation of the
odd-sector variables ($C, Q, W$), as seen from Eqs.~(\ref{gaugetrans}). In
fact all these variables transform only in terms of $\eta$, which has been
fixed such that $C$ could be eliminated. Then the other two odd-sector
variables stay arbitrary, unaffected by the three other gauge choices.

Finally, we comment on the elimination of the even-sector variable $\delta 
\mathcal{N}$. By doing so, the interpretation of the Lagrangian variables in
terms of the geometric quantities defined in the 2+1+1 formalism continues
to hold even in the presence of perturbations. Such a condition is
equivalent to imposing hypersurface-orthogonality of the vector field $l^{a}$%
. The last requirement could be relaxed such that the vector $l^{a}$
acquires vorticity at a perturbative level. However, this would imply to
develop a more involved formalism, allowing at least for a new scalar, a new
vectorial and a new tensorial degree of freedom (and all the scalars formed
from them). Then we can choose another gauge $\widetilde{P}=0$, as commonly
used in past works. Such a generalization of the formalism for the
even-parity perturbations is left for a subsequent work.

\section{Odd-mode perturbation dynamics}
\label{perturbsec} 

We proceed with the analysis of the odd-parity perturbations by expanding
the action up to second order to discuss the dynamical evolution 
of them.

\subsection{Second-order perturbed Lagrangian\label{L2}}

We expand the action (\ref{action}) at second order for the odd-type
perturbations in order to derive linear perturbation equations of motion. As
the even and odd sectors decouple in the second-order perturbed Lagrangian,
at a formal level, we could just switch off all even-type variables as 
\begin{equation}
P=V=\delta N=\delta M=\zeta =0\,.
\end{equation}
Then the second-order contribution to the Lagrangian density for the odd
modes is given by 
\begin{equation}
\delta _{2}\mathscr{L}^{\mathrm{odd}}=\bar{L}^{\mathrm{EFT}}_{0}\delta _{2}%
\sqrt{-g}+\delta \sqrt{-g}\delta L^{\mathrm{EFT}}+\sqrt{-\bar{g}}\,\delta
_{2}L^{\mathrm{EFT}}\,,  \label{Lagden2}
\end{equation}
where $\delta _{2}$ represents second-order variations.

The second-order contribution to the line element reads 
\begin{equation}
\delta _{2}\left( ds^{2}\right) =(\delta N_{a}\delta N^{a}-\delta
N^{2})dt^{2}+2\delta N_{a}\delta M^{a}dtd{r}+\left( \delta M_{a}\delta
M^{a}+\delta M^{2}\right) d{r}^{2}+2\zeta ^{2}\bar{h}_{ab}dx^{a}dx^{b}\;.
\label{metric2}
\end{equation}
By employing Eqs.~(\ref{metric1}) and (\ref{metric2}), it follows that 
\begin{equation}
\delta _{2}\sqrt{-g}=\frac{\sqrt{-\bar{g}}}{2}\left[ \bar{g}^{ab}\delta
_{2}g_{ab}+\frac{1}{4}\left( \bar{g}^{ab}\bar{g}^{cd}-2\bar{g}^{ac}\bar{g}
^{bd}\right) \delta g_{ab}\delta g_{cd}\right] =0\,.  \label{detg2}
\end{equation}
Thus the first term on the rhs of Eq.~(\ref{Lagden2}) vanishes identically.
Similarly the second term on the rhs of Eq.~(\ref{Lagden2}) vanishes, since
by virtue of Eq.~(\ref{delg}) the first-order variation $\delta \sqrt{-g}$
consists only of even-mode contributions.

Next we expand the Lagrangian up to second order. Before doing so, we note
that the linear and quadratic perturbations of $L$, $\mathcal{L}$, $K$, 
$\mathcal{K}$ and $\mathcal{R}$ arise from even modes only 
[see Eqs.~(\ref{calLexp}), (\ref{Ks}) and (\ref{Rperturb})], so they do not contribute to
the odd-mode dynamics. As a result, the second-order Lagrangian for the
odd-type perturbations becomes extremely simple 
(depending on 4 variables only out of 11): 
\begin{equation}
\delta _{2}L^{\mathrm{EFT}}=L_{\mathfrak{M}}^{\mathrm{EFT}}\delta _{2}%
\mathfrak{M} +L_{\mathfrak{K}}^{\mathrm{EFT}}\delta _{2}\mathfrak{K}
+L_{\varkappa }^{\mathrm{EFT}}\delta _{2}\varkappa +L_{\lambda }^{\mathrm{EFT%
}}\delta _{2}\lambda \,.  \label{del2Lag}
\end{equation}

Substituting Eqs.~(\ref{Nadec}) and (\ref{Madec}) into Eqs.~(\ref{calLexp})
and (\ref{Ks}), then integrating by parts (employing once again the
generalized Stokes theorem for manifolds without boundaries), the
second-order factors in $\delta _{2}L^{\mathrm{EFT}}$ can be explicitly
expressed in terms of the odd-type variables: 
\begin{eqnarray}
&&\delta _{2}\mathfrak{M}=\left( \bar{D}W\right) ^{2}\,,\quad \delta
_{2}\lambda =\frac{1}{2\bar{M}}\left[ \left( \bar{D}^{2}W\right) ^{2}-\frac{2%
}{{r}^{2}}\left( \bar{D}W\right) ^{2}\right] \,,\quad \delta _{2}\varkappa =%
\frac{1}{2\bar{N}^{2}}\left[ \left( \bar{D}^{2}Q\right) ^{2}-\frac{2}{{r}^{2}%
}\left( \bar{D}Q\right) ^{2}\right] \,,  \notag \\
&&\delta _{2}\mathfrak{K}=\frac{1}{4\bar{N}^{2}\bar{M}^{2}}\left[ \left( 
\bar{D}\dot{W}\right) ^{2}+\left( \bar{D}Q^{\prime }\right) ^{2}-2\bar{D}^{a}%
\dot{W}\bar{D}_{a}Q^{\prime }+\frac{4}{{r}}\left( \bar{D}^{a}\dot{W}\bar{D}%
_{a}Q-\bar{D}^{a}Q\bar{D}_{a}Q^{\prime }\right) +\frac{4}{{r}^{2}}\left( 
\bar{D}Q\right) ^{2}\right] \,,  \label{delta2}
\end{eqnarray}
where the notations $\bar{D}^{2}\equiv \bar{D}^{a}\bar{D}_{a}$ and $\left( 
\bar{D}f\right) ^{2}\equiv \bar{D}^{a}f\bar{D}_{a}f$ have been introduced
for $f\equiv \left( Q,W\right) $.

Substituting Eqs.~(\ref{detg2})-(\ref{delta2}) and $\delta \sqrt{-g}=0$ into
the second-order Lagrangian density (\ref{Lagden2}) for the odd modes, we
finally obtain 
\begin{equation}
\delta _{2}\mathscr{L}^{\mathrm{odd}}=\sqrt{-\bar{g}}\left\{ a_{1}\left( 
\bar{D}\dot{W}-\bar{D}Q^{\prime }+\frac{2}{{r}}\bar{D}Q\right) ^{2}
+a_{2}\left[ \left( \bar{D}^{2}Q\right) ^{2}-\frac{2}{{r}^{2}}\left( 
\bar{D}Q\right) ^{2}\right] +{a_{3}}\left( \bar{D}^{2}W\right) ^{2}+a_{4}\left( 
\bar{D}W\right) ^{2}\right\} \,,  
\label{quadLagden}
\end{equation}
where the coefficients $a_{i}$ ($i=1,\cdots,4$) are 
\begin{equation}
a_{1}=\frac{L_{\mathfrak{K}}^{\mathrm{EFT}}}{4\bar{N}^{2}\bar{M}^{2}}
\,,\qquad a_{2}=\frac{L_{\varkappa }^{\mathrm{EFT}}}{2\bar{N}^{2}}\,,\qquad
a_{3}=\frac{L_{\lambda }^{\mathrm{EFT}}}{2\bar{M}^{2}}\,,\qquad a_{4}
=L_{\mathfrak{M}}^{\mathrm{EFT}}-\frac{2}{r^{2}}a_{3}\,.  
\label{coeffs}
\end{equation}
{}From the second-order Lagrangian density (\ref{quadLagden}), we will
derive the equations of motion for the odd-sector perturbations in the
next subsection. We remark that the Lagrangian density 
(\ref{quadLagden}) is quadratic in the odd-mode perturbations $Q$ and $W$, 
so in what follows
we will refer to this Lagrangian contribution as quadratic.

\subsection{Perturbation equations in the harmonics expansion}
\label{harsec}

We rewrite the quadratic action 
$S_2=\int d^{4}x\,\delta _{2}\mathscr{L}^{\mathrm{odd}}$ 
in the following form
\begin{equation}
\delta _{2}\mathscr{L}^{\mathrm{odd}}=\sqrt{-\bar{g}}\left[ -a_{1}\left( 
\dot{W}-Q^{\prime }+\frac{2}{{r}}Q\right) \bar{D}^{2}\left( \dot{W}
-Q^{\prime }+\frac{2}{{r}}Q\right) +a_{2}Q\bar{D}^{2}\left( \bar{D}^{2}
+\frac{2}{{r}^{2}}\right) Q+W\bar{D}^{2}\left( 
{a_{3}}\bar{D}^{2}-a_{4}\right) W\right] \,,  \label{Lquad1}
\end{equation}
in which we have dropped covariant total divergence terms. 
The resulting equations of motion derived by varying 
$W$ and $Q$ are given, respectively, by 
\begin{equation}
\bar{D}^{2} \Psi^{(1)}=0\,,\qquad
\Psi^{(1)} \equiv a_{1}\frac{\partial }{\partial t}\left( \dot{W}-Q^{\prime
}+\frac{2Q}{{r}}\right) +\left( {a_{3}}\bar{D}^{2}-a_{4}\right) W\,,  \label{4newa}
\end{equation}
and 
\begin{equation}
\frac{1}{\sqrt{-\bar{g}}r^{2}}\frac{\partial }{\partial r}\left[ \sqrt{
-\bar{g}}a_{1}r^{2}\bar{D}^{2}\left( \dot{W}-Q^{\prime }+\frac{2}{{r}}Q\right) 
\right] -a_{2}\bar{D}^{2}\left( \bar{D}^{2}+\frac{2}{{r}^{2}}\right) Q =0\,.  
\label{4newaux}
\end{equation}

Since $\sqrt{-\bar{g}}=\bar{N}\bar{M}\sqrt{\bar{h}}=\bar{N}\bar{M}r^{2}\sin
\theta $ and $\bar{D}_{a}$ is the covariant derivative compatible with
the metric \thinspace $h_{ab}$, it follows that $\bar{D}_{a}\sqrt{-\bar{g}}=0$. 
On using this identity and the fact that $r^{2}\bar{D}^{2}$ has no radial
dependence (i.e., it commutes with $\partial /\partial r$), 
Eq.~(\ref{4newaux}) reads
\begin{equation}
\bar{D}^{2} \Psi^{(2)}=0\,,\qquad 
\Psi^{(2)} \equiv
\frac{1}{\sqrt{-\bar{g}}}\frac{\partial }{\partial r}\left[
\sqrt{-\bar{g}}a_{1}\left( \dot{W}-Q^{\prime }+\frac{2}{{r}}Q\right) \right]
-a_{2}\left( \bar{D}^{2}+\frac{2}{{r}^{2}}\right) Q \,.
\label{4newb}
\end{equation}
Hence Eqs.~(\ref{4newa}) and (\ref{4newb}) are of the form 
$\bar{D}^{2}\Psi^{(i)}=0$ with $i=1,2$.
These are fourth-order coupled differential equations, but 
in the expressions of $\Psi^{(i)}$ they contain time 
and radial derivatives up to second orders alone. 

In the following, we expand the angular part of the odd-mode 
perturbations $f\equiv \left( Q,W\right)$ 
in terms of spherical harmonics, i.e.,
\begin{equation}
f(t,r,\theta ,\varphi )=\sum_{l,m}f_{lm}(t,r)Y_{l}^{m}\,.
\end{equation}
A similar decomposition of the differential expressions 
$\Psi^{(i)}$ ($i=1,2$) is given by 
\begin{equation}
\Psi^{(i)} (t,r,\theta ,\varphi )=\sum_{l,m}\Psi^{(i)}_{lm}(t,r)Y_{l}^{m}\,.
\end{equation}
Each mode obeys the identity 
\begin{equation}
r^{2}\bar{D}^{2}\left[ \Psi^{(i)}_{lm}(t,r)Y_{l}^{m}\right] +l\left( l+1\right) 
\left[ \Psi^{(i)}_{lm}(t,r)Y_{l}^{m}\right] =0\,.  \label{idsphe}
\end{equation}

The differential order of Eqs.~(\ref{4newa}) and (\ref{4newb}) 
can be reduced by two, i.e., 
\begin{equation}
\sum_{l,m}l\left( l+1\right) \Psi^{(i)}_{lm}(t,r)Y_{l}^{m}=0\,, \qquad(i=1,2),
\label{eqsdecomposed}
\end{equation}
or explicitly 
\begin{eqnarray}
& &\sum_{l}l\left( l+1\right) \Psi^{(1)}_l=0\,,\qquad
\Psi^{(1)}_l \equiv 
{a_{1}}\,\frac{\partial }{\partial t}
\left( \dot{W}_{l}-Q_{l}^{\prime }+\frac{2}{r}Q_{l}\right) -\left[ 
a_{3}\frac{l\left( l+1\right) }{r^{2}}+a_{4}\right] W_{l} \,,
\label{eomW} \\
& &
\sum_{l}l\left( l+1\right) \Psi^{(2)}_l=0\,,\qquad
\Psi^{(2)}_l \equiv 
\frac{1}{\sqrt{-\bar{g}}}\,\frac{\partial 
}{\partial r}\left[ \sqrt{-\bar{g}}\,a_{1}\left( \dot{W}_{l}-Q_{l}^{\prime }+
\frac{2}{r}Q_{l}\right) \right] +{a_{2}}\frac{l\left( l+1\right) -2}{r^{2}}
Q_{l}\,.
\label{eomQ}
\end{eqnarray}
Note that we have introduced the notations $f_{l}\equiv \sum_{m}f_{lm}Y_{l}^{m}$.
The $f_{l}$ modes are orthogonal to each other due to the orthogonality 
of spherical harmonics, so that $\Psi_l^{(1)}$ and $\Psi_l^{(2)}$ 
vanish for $l \neq 0$.
Hence we have derived a sequence of second-order differential 
equations $\Psi_l^{(i)}=0$ ($i=1,2$)
holding for each non-zero $l$.

There exists a second time derivative of $W_l$ in Eq.~(\ref{eomW}), 
so this corresponds to a dynamical equation of motion for $W_l$.
The variable $Q_{l}$ appears only algebraically in 
the second-order Lagrangian density (\ref{quadLagden}) 
and through a first temporal derivative in Eq.~(\ref{eomW}).
Since Eq.~(\ref{eomQ}) contains only a first time derivative 
of $W_l$ with no time derivatives of $Q_l$,  
this is a constraint equation in the Lagrangian sense.
In Sec.~\ref{degreesec} we shall address the issue of 
a true dynamical degree of freedom for general $l$ 
by using a method of the Lagrange multiplier.
Before doing so, we shall discuss the specific cases 
of $l=0, 1$ in the next subsection.

\subsection{Monopolar and dipolar perturbations}

\subsubsection{Monoploar mode $(l=0)$}

The monopolar perturbations trivially
obey Eqs.~(\ref{eomW})-(\ref{eomQ}), so they do not contribute to the
dynamics. In fact, after integrations by parts, the quadratic odd-mode
Lagrangian density (\ref{quadLagden}) can be written in a form containing
exclusively Laplacian terms: 
\begin{eqnarray}
\delta _{2}\mathscr{L}^{\mathrm{odd}} &=&\sqrt{-\bar{g}}\Bigg[-a_{1}\left( 
\dot{W}-Q^{\prime }+\frac{2}{{r}}Q\right) \left( \bar{D}^{2}\dot{W}
-\bar{D}
^{2}Q^{\prime }+\frac{2}{{r}}\bar{D}^{2}Q\right) +a_{2}\left( 
\bar{D}^{2}Q\right)\left(\bar{D}^{2}+\frac{2}{{r}^{2}}\right)Q 
\notag \\
&&~~~~~~~+\left( \bar{D}^{2}W\right) \left(a_3\D^2-a_{4}\right)W\Bigg]\,,
\end{eqnarray}
all of which identically vanish for $l=0$. 
In the following we consider only perturbations without a monopolar
contribution.

\subsubsection{Dipolar mode $(l=1)$}

For the dipolar perturbations, the last term of Eq.~(\ref{hdec}), which 
contains the term $C$, vanishes due to the identity (\ref{idsphe}). 
Hence there is no need to eliminate $C$ by gauge fixing, 
so that the respective gauge degree of freedom can be used up as 
\begin{equation}
\eta =-r^{2}\int dr\frac{W_{1}}{r^{2}}+r^{2}\mathcal{C}_{0}(t,\theta,
\varphi )\,,  \label{digauge}
\end{equation}
where $\mathcal{C}_{0}(t,\theta ,\varphi )$ is an integration function. 
With this choice, $\widetilde{W}_{1}=0$ and $\widetilde{Q}_{1}=Q_{1}+r^{2}
\dot{\mathcal{C}}_{0}(t,\theta ,\varphi )$. 
Omitting tildes as before and noting that the last term of 
Eq.~(\ref{eomQ}) also vanishes due to the identity 
(\ref{idsphe}), Eqs.~(\ref{eomW})-(\ref{eomQ}) is 
simplified as 
\begin{eqnarray}
&&\frac{\partial }{\partial t}\left( Q_{1}^{\prime }-\frac{2}{r}Q_{1}\right)
=0\,,  \label{d1} \\
&&\frac{\partial }{\partial r}\left[ \sqrt{-\bar{g}}\,a_{1}\left(
Q_{1}^{\prime }-\frac{2}{r}Q_{1}\right) \right] =0\,.  \label{d2}
\end{eqnarray}

The dynamical degree of freedom $W$ does not appear in Eqs.~(\ref{d1})-(\ref{d2}), 
suggesting that dipolar perturbations are non-dynamical.
Indeed, direct integration of Eqs.~(\ref{d1})-(\ref{d2}) leads to 
\begin{equation}
Q_{1}=r^{2}\mathcal{C}_{1}(\theta ,\varphi )\int \frac{dr}{\sqrt{-\bar{g}}
a_{1}r^{2}}+r^{2}\mathcal{C}_{2}(t,\theta ,\varphi )\,,  \label{diQ}
\end{equation}
where $\mathcal{C}_{1,2}$ are integration functions. The remaining gauge
degree of freedom can be exploited as $\dot{\mathcal{C}}_{0}=-\mathcal{C}_{2}$, 
so the time dependence is completely eliminated from the
dipolar odd-mode perturbations. As discussed in Ref.~\cite{Zerilli}, the 
time-independent contribution to $Q_{1}$ appearing as the first term 
on the r.h.s. of Eq.~(\ref{diQ}) is related to the
angular momentum induced by the dipolar perturbation.

\subsection{Dynamical degree of freedom for $l \geq 2$}
\label{degreesec}

The Lagrangian density (\ref{Lquad1}) possesses first and second derivatives, 
which appear quadratically. Hence some of the terms 
would be of fourth order in spatial derivatives by partial integration 
(while the time derivatives remain of second order). 
This is why the perturbation Eqs.~(\ref{4newa}) and (\ref{4newb}) 
involve fourth-order spatial 
differentiations. For $l \geq 2$ these equations of motion reduce to 
the form $\Psi_l^{(1)}=0$ and $\Psi_l^{(2)}=0$ under the expansion 
of spherical harmonics, 
where $\Psi_l^{(i)}$ ($i=1,2$) are given by 
Eqs.~(\ref{eomW}) and (\ref{eomQ}).

As we already mentioned in Sec.~\ref{harsec}, the first equation ($\Psi_l^{(1)}=0$) 
describes the dynamical evolution of the variable $W_l$, whereas the second 
one ($\Psi_l^{(2)}=0$) corresponds to a constraint involving a second spatial 
derivative of the field $Q_l$. Since the latter constraint equation is not 
directly solved for $Q_l$, it is difficult to derive a closed-form differential 
equation for $W_l$ by eliminating the $Q_l$-dependent terms appearing in the 
equation $\Psi_l^{(1)}=0$. 
This obstacle can be circumvented by using the method of a Lagrange 
multiplier. In fact, this method was employed to study the linear perturbations 
on a spherically symmetric background in modified Gauss-Bonnet gravity \cite{LagMult} 
and it was further applied to Horndeski theory \cite{suyama}.

Introducing the Lagrange multiplier vector $Y^{a}$, the Lagrangian density 
equivalent to Eq.~(\ref{quadLagden}) is given by 
\begin{equation}
\delta _{2}\mathscr{L}^{\mathrm{odd}}=\sqrt{-\bar{g}}\left\{ a_{1}\left[
2Y^{a} \bar{D}_{a} \left( \dot{W}-Q^{\prime}+\frac{2}{r}Q \right) 
-Y^{2}\right] +a_{2}\left[ \left( \bar{D}^{2}Q\right) ^{2}-
\frac{2}{{r}^{2}}\left( \bar{D}Q\right) ^{2}\right] +{a_{3}}\left( \bar{D}
^{2}W\right) ^{2}+a_{4}\left( \bar{D}W\right) ^{2}\right\} \,,
\label{Lquad2}
\end{equation}
where $Y^2=Y^{a}Y_{a}$. 
Variation of Eq.~(\ref{Lquad2}) with respect to $Y^a$ leads 
to $Y_{a}=\bar{D}_{a} [ \dot{W}-Q^{\prime }+(2/r)Q]$. 
Substituting this relation into Eq.~(\ref{Lquad2}), we recover 
the original Lagrangian density (\ref{quadLagden}).
 
Defining the Lagrange multiplier potential $Z$ as 
$Y^{a}=\bar{D}^{a}Z$, the Lagrangian density (\ref{Lquad2}) 
is characterized by two scalar fields 
$W$ and $Q$ plus the auxiliary scalar field $Z$. 
Varying Eq.~(\ref{Lquad2}) in terms of $W$ and $Q$, 
we obtain 
\begin{eqnarray}
&&\bar{D}^{2}\left[ a_{1}\dot{Z}+\left( {a_{3}}\bar{D}^{2}-a_{4}\right) W
\right]  =0\,,  \label{eomW3} \\
&&\bar{D}^{2}\left[ \frac{1}{\sqrt{-\bar{g}}}
\frac{\partial }{\partial r}\left( \sqrt{-\bar{g}}
a_{1}Z\right) -a_{2}\left( \bar{D}^{2}+\frac{2}{{r}^{2}}
\right) Q\right]  =0\,.  \label{eomQ3}
\end{eqnarray}
For $l\geq 2$ the $\bar{D}^{2}$ operators acting on the square
brackets can be formally omitted, so the $l$-th multipolar components 
$W_l$ and $Q_l$ obey the following equations:
\begin{eqnarray}
W_{l} &=&\frac{a_{1}r^2}{a_{3}l\left( l+1\right) +a_{4}r^{2}}\,
\dot{Z}_{l}\,, \label{Wlm} \\
Q_{l} &=&-\frac{r^{2}}{a_2 \left( l+2 \right) \left( l-1 \right) \sqrt{-\bar{g}}}\,
\frac{\partial }{\partial r}\left( \sqrt{-\bar{g}}\,a_{1}Z_{l}
\right) \,, \label{Qlm}
\end{eqnarray}
where $Z_l$ is the $l$-th component of $Z$.

Equations (\ref{Wlm}) and (\ref{Qlm}) show that both $W_l$ and $Q_l$ 
are directly known from $Z_l$. On using the last of Eq.~(\ref{coeffs}), 
we can also write Eq.~(\ref{Wlm}) of the form
\begin{equation}
\allowbreak W_{l}=\frac{r^{2}}{a_3\left( l+2\right) \left( l-1\right)}
\left( a_{1}\dot{Z}_{l}-L_{\mathfrak{M}}^{\mathrm{EFT}}W_{l}\right)\,.
\label{Wlm1}
\end{equation}
Substituting Eqs.~(\ref{Qlm}) and (\ref{Wlm1}) into the $l$-th
component of the multipolar decomposition of the Lagrangian density 
(\ref{Lquad2}), using $Y^{a}=\bar{D}^{a}Z$, and integrating it by parts, 
we finally obtain
\begin{equation}
\delta _{2}\mathscr{L}_{l}^{\mathrm{odd}}=
\frac{l( l+1)}{( l+2)( l-1)}\sqrt{-\bar{g}}
\left[ -\frac{a_{1}^2}{a_{3}
}\dot{Z}_{l}^{2}-\frac{a_{1}^2}{a_{2}}Z_{l}^{\prime 2}
-a_1 (\bar{D} Z_l)^2
-U^{\mathrm{H}}(r)Z_{l}^{2}
+\frac{a_1}{a_3}L_{\mathfrak{M}}^{\mathrm{EFT}}
W_{l}\dot{Z}_{l}\right]\,, 
\label{quad5}
\end{equation}
where the potential $U^{\mathrm{H}}(r)$ is given by
\begin{equation}
U^{\mathrm{H}}(r)=-a_1\frac{\partial}{\partial r}\left[ \frac{1}{\sqrt{-\bar{g}
}a_{2}}\frac{\partial }{\partial r}\left( \sqrt{-\bar{g}}a_{1}\right) \right]
-\frac{2a_1}{r^{2}}\,,  \label{UH}
\end{equation}
or more explicitly, 
\begin{equation}
U^{\mathrm{H}}(r)=-\frac{a_{1}^2}{a_{2}}\left[ \frac{\bar{N}^{\prime \prime }}{
\bar{N}}+\frac{\bar{M}^{\prime \prime }}{\bar{M}}-\frac{\bar{N}^{\prime 2}}{
\bar{N}^{2}}-\frac{\bar{M}^{\prime 2}}{\bar{M}^{2}}-\frac{2}{r^{2}}+\left( 
\frac{a_{1}^{\prime }}{a_{1}}-\frac{a_{2}^{\prime }}{a_{2}}\right) \left( 
\frac{\bar{N}^{\prime }}{\bar{N}}+\frac{\bar{M}^{\prime }}{\bar{M}}+\frac{2}
{r}\right) +\frac{a_{1}^{\prime \prime }}{a_{1}}-\frac{a_{1}^{\prime
}a_{2}^{\prime }}{a_{1}a_{2}}\right] -\frac{2a_1}{r^{2}}\,. \label{UHdet}
\end{equation}
The superscript in $U^{\mathrm{H}}(r)$ has been introduced to point out that
in the Horndeski limit it reduces to the potential (24) of Ref.~\cite{suyama}. 
Using the relation $\bar{D}^2 Z_l=-l(l+1)Z_l/r^2$, the third term 
in the square bracket of Eq.~(\ref{quad5}) is 
equivalent to $-a_1 l(l+1)Z_l^2/r^2$ up to a boundary term.

The last term in the square bracket of Eq.~(\ref{quad5}) gives rise to 
a contribution $\dot{Z}_l^2$ with a coefficient including the multipolar 
index $l$ by virtue of Eq.~(\ref{Wlm}). 
In this case the propagation speeds are different for each multipolar mode, 
so the global interpretation of the perturbation $Z_l$ and its propagation 
speeds become far from trivial. 
Hence, in the following, we impose the following condition 
\begin{equation}
L_{\mathfrak{M}}^{\mathrm{EFT}}=0\,.  
\label{cond1}
\end{equation}
In fact, this is satisfied both in the Horndeski theory and in
the GLPV theory ($i=2,3,4$ for our cases of interest). 

Under the condition (\ref{cond1}) the second-order Lagrangian 
density is expressed solely by the quantity $Z_l$ and its time and 
spatial derivatives, in a mode-independent way.  
As a result, $Z_l$ is a master variable
governing the dynamics of the odd-mode perturbations.
Comparing Eqs.~(\ref{eomW3}) and (\ref{eomQ3}) with 
Eqs.~(\ref{4newa}) and (\ref{4newb}), respectively, 
there is the correspondence $Z \to \dot{W}-Q'+2Q/r$, which 
also arises by varying the Lagrangian density 
(\ref{Lquad2}) for the Lagrange multiplier potential $Z$.  
While $Q$ and $W$ were
eliminated from the Lagrangian density by their respective equations of
motion, Eqs.~(\ref{Wlm}) and (\ref{Qlm}), we stress that this third equation
of motion $Z=\dot{W}-Q^{\prime }+2Q/r$ was \textit{not} exploited for
deriving Eq.~(\ref{quad5}). 
In fact, after the substitution of Eqs.~(\ref{Wlm}) and (\ref{Qlm}), 
the Lagrangian density (\ref{quad5}) already contains the
dynamics of the third field $Z$.
If we were to make the additional substitution $Z_l \to \dot{W}_l-Q'_l+2Q_l/r$,
the Lagrangian density (\ref{Lquad2}) would reduce to a boundary term 
$\delta _{2}\mathscr{L}^{\mathrm{odd}}=-(\partial/\partial r)
(\sqrt{-\bar{g}}a_1 Q l(l+1) Z_l/r^2)$, which is irrelevant to the true dynamics 
of perturbations.

On using the equations of motion following from the variation 
of Eq.~(\ref{quad5}) with respect to $Z_l$, we can discuss the 
stability of the odd-type perturbations. In the next section 
we shall address this issue.


\section{No-ghost conditions and avoidance of Laplacian instabilities}
\label{stabsec} 

In the previous section we have seen that in an expansion with respect to 
spherical harmonics there is no monopolar contribution to the odd modes 
and the dipolar mode is non-dynamical. 
In the following we proceed with the stability analysis
of quadrupolar and higher multipolar contributions to the odd-mode
perturbations ($l \geq 2$), governed by the quadratic Lagrangian density 
(\ref{quad5}) under the condition (\ref{cond1}). 

\subsection{Generalized Horndeski class}

We categorize theories satisfying the condition (\ref{cond1}) as the
generalized Horndeski class (including the GLPV theory). In this case the
quadratic Lagrangian could depend on the odd-mode variable $W$, which
generates the odd-mode contribution to $\delta M_{a}$ through 
Eq.~(\ref{Madec}). 
Nevertheless, the Lagrangian for the background dynamics does not
depend on the particular combination $\mathfrak{M}\equiv M_{a}M^{a}$.
Whenever Eq.~(\ref{cond1}) holds, the quadratic Lagrangian density 
(\ref{quad5}) leads to a second-order differential equation for the decoupled
master variable $Z_l$ and the usual stability conditions can be imposed 
on this equation.

The condition for avoidance of the scalar ghost (no negative kinetic
term) is satisfied for $a_{3}<0$, i.e., 
\begin{equation}
L_{\lambda }^{\mathrm{EFT}}<0\,.  \label{noghostQW}
\end{equation}
For the modes with the large wave numbers along the radial or tangential
directions, many terms of Eq.~(\ref{quad5}) are suppressed. In particular, 
$U^{\mathrm{H}}(r)$ as well as the third (for radial modes) or second (for
tangential modes) terms are sub-dominant in the high-frequency limit. In
these two regimes the dispersion relations following from the Lagrangian
density (\ref{quad5}) are given, respectively, by 
\begin{equation}
\omega^2+\frac{a_{3}}{a_{2}}k_{r}^2=0\,, \qquad 
\omega^2 +\frac{a_{3}}{a_{1}}
k_{\Omega}^2=0\,,
\end{equation}
where $\omega$ is the angular frequency, $k_r$ and $k_{\Omega}$ are the wave
numbers along the radial and tangential directions respectively. 
Introducing proper time $\tau=\int \bar{N} dt$ and tortoise coordinate 
$r_{\ast}=\int \bar{M} dr$, the squared sound speeds of fluctuations 
along the radial and tangential directions read 
\begin{equation}
c_{r}^{2}\equiv \frac{\bar{M}^{2}k_r^2}{\bar{N}^2\omega^2}= -\frac{\bar{M}%
^{2}a_{3}}{\bar{N}^{2}a_{2}}=-\frac{L_{\lambda }^{\mathrm{EFT}}} {%
L_{\varkappa }^{\mathrm{EFT}}}\,,\qquad c_{\Omega }^{2}\equiv \frac{%
k_{\Omega}^2}{\bar{N}^2\omega^2}=-\frac{a_{3}}{\bar{N}^{2}a_{1}}= -\frac{%
2L_{\lambda }^{\mathrm{EFT}}}{L_{\mathfrak{K}}^{\mathrm{EFT}}}\,,
\end{equation}
respectively. Under the no-ghost requirement (\ref{noghostQW}), the
conditions for the absence of Laplacian instabilities, i.e., $c_{r}^{2}>0$
and $c_{\Omega}^{2}>0$, take a remarkably simple form 
\begin{equation}
L_{\varkappa }^{\mathrm{EFT}}>0\,,\qquad L_{\mathfrak{K}}^{\mathrm{EFT}}>0\,.
\label{nolapQW}
\end{equation}

These simple stability conditions acquire geometrical significance, as 
$\mathfrak{K}$ is the length squared of the normal fundamental vector, while 
$\varkappa $ and $\lambda $ are the traces of the squares of the two
extrinsic curvature tensors of the spheres. These are quantities appearing
in the 2+1+1 decomposition of the covariant derivatives of the two normal
vectors to the spheres, Eqs.~(\ref{nablan})-(\ref{nablal}). The additional
quantities of these decompositions are the normal fundamental scalars and
accelerations. They however do not contribute to the stability conditions
for the odd modes as the normal fundamental scalars $\mathcal{L}$ and $%
\mathcal{K}$ are even-mode variables, while the accelerations $\alpha ^{a}$
and $\beta ^{a}$ appear in the action only through the curvature scalar 
$\mathcal{R}$ under divergences. Hence their rotation-free part alone
survives under the Helmholtz decomposition, which again generates the even
modes.

The stability conditions (\ref{noghostQW}) and (\ref{nolapQW}) can be
further specified for the particular case of the Horndeski theory with 
$L_{2,3,4}^{\mathrm{H}}$ and the GLPV theory with 
$L_{2,3,4}^{\mathrm{GLPV}}$ discussed in Sec.~\ref{hornsec}. 
For this we first remark that, according to Eqs.~(\ref{L4adm}) and (\ref{exhornL4}), 
only the contributions $L_{4}^{\mathrm{H}}$ and $L_{4}^{\mathrm{GLPV}}$ 
depend on the variables $\lambda $, $\varkappa$ and $\mathfrak{K}$.

In the Horndeski theory, the stability conditions (\ref{noghostQW}) and 
(\ref{nolapQW}) read 
\begin{equation}
-L_{\lambda }^{\mathrm{H}}=\frac{1}{2}L_{\mathfrak{K}}^{\mathrm{H}}
=G_{4}-2XG_{4X}>0\,,\qquad L_{\varkappa }^{\mathrm{H}}=G_{4}>0\,\,.
\label{cond}
\end{equation}
The first of these conditions exactly corresponds to Eq.~(25) or Eq.~(28) of
Ref.~\cite{suyama} (these two conditions coincide when 
$L_{5}^{\mathrm{H}}=0$). 
The second is the condition imposed in Ref.~\cite{suyama} for avoiding
gradient instabilities, when $L_{5}^{\mathrm{H}}=0$. Since $X>0$, the first
condition (\ref{cond}) gives information beyond the second one
only for $G_{4X}>0 $.

In the GLPV theory, the stability conditions reduce to 
\begin{equation}
L_{\lambda }^{\mathrm{GLPV}}=-\frac{1}{2} L_{\mathfrak{K}}^{\mathrm{GLPV}
}=A_{4}<0\,,\qquad L_{\varkappa }^{\mathrm{GLPV}}=B_{4}>0\,.  
\label{cond2}
\end{equation}
It is easy to see that, in the Horndeski limit characterized by 
Eq.~(\ref{H1}), these reduce to Eq.~(\ref{cond}).

\subsection{Stability conditions for covariant Galileon models}

\subsubsection{Covariantized Galileons}

The original Galileon model advocated in Ref.~\cite{Nicolis} is composed of
five Lagrangians invariant under the Galilean symmetry $\partial_{\mu}\phi
\to\partial_{\mu}\phi+b_{\mu}$ in the Minkowski background. The equations of
motion remain of second order by virtue of this symmetry. In the curved
background, the original Galileon model can be covariantized by replacing
coordinate derivatives with covariant derivatives. This ``covariantized
Galileon'' belongs to a particular case of the GLPV theory given by the
Lagrangians (\ref{exhornLa2})-(\ref{exhornLa4}) with the functions 
\begin{equation}
A_{2}=c_{2}X\,,\qquad A_{3}=c_{3}X^{3/2}\,,\qquad 
A_{4}=-\frac{M_{\mathrm{pl}
}^{2}}{2}-c_{4}X^{2}\,,\qquad B_{4}=\frac{M_{\mathrm{pl}}^{2}}{2}\,,
\label{covGal}
\end{equation}
where $c_{2,3,4}$ are constants. Here we have taken into account the
Einstein-Hilbert term $M_{\mathrm{pl}}^2R/2$ in the Lagrangian, 
where $M_{\mathrm{pl}}$ is the reduced Planck mass.

In general space-time, the theory described by (\ref{covGal}) contains
derivatives higher than second order. On the flat isotropic cosmological
background, however, the equations of motion for the background and linear
perturbations are second order without a new propagating degree of freedom 
\cite{Gleyzes}. This result was obtained by considering the constant-time
hypersurfaces, such that the scalar field plays the role of time. A similar
argument may also work for the spherically symmetric background due to the
high degree of symmetry, in which case the scalar field takes the role of a
radial coordinate $r$. In fact, substituting Eq.~(\ref{covGal}) into the
background equations of motion (\ref{GLPVeom00})-(\ref{GLPVeom22}), we
obtain 
\begin{eqnarray}
&&\frac{M_{\mathrm{pl}}^{2}}{r}\left( \frac{1}{r}-\frac{1}{\bar{M}^{2}r}+ 
\frac{2\bar{M}^{\prime }}{\bar{M}^{3}}\right) +{c}_{2}X+\frac{3{c}_{3}X} 
{\bar{M}^{2}}\left( \frac{\phi ^{\prime }\bar{M}^{\prime }}{\bar{M}}-\phi
^{\prime \prime }\right) -\frac{2{c}_{4}X}{\bar{M}^{2}r}\left( \frac{X}{r}- 
\frac{10X\bar{M}^{\prime }}{\bar{M}}+\frac{8\phi ^{\prime }\phi ^{\prime
\prime }}{\bar{M}^{2}}\right) =0\,,  \notag  \label{GLeq1} \\
&&\frac{M_{\mathrm{pl}}^{2}}{r}\left( \frac{1}{r}-\frac{1}{\bar{M}^{2}r}- 
\frac{2\bar{N}^{\prime }}{\bar{M}^{2}\bar{N}}\right) -{c}_{2}X-\frac{3{c}
_{3}X\phi ^{\prime }}{\bar{M}^{2}}\left( \frac{2}{r}+\frac{\bar{N}^{\prime} 
}{\bar{N}}\right) -\frac{10{c}_{4}X^{2}}{r\bar{M}^{2}}\left( \frac{1}{r}+ 
\frac{2\bar{N}^{\prime }}{\bar{N}}\right) =0\,,  \notag \\
&&\frac{M_{\mathrm{pl}}^{2}}{\bar{M}^{2}}\left[ \frac{\bar{M}^{\prime }} {%
\bar{M}r}-\frac{\bar{N}^{\prime \prime }}{\bar{N}}-\frac{\bar{N}^{\prime}}{ 
\bar{N}}\left( \frac{1}{r}-\frac{\bar{M}^{\prime }}{\bar{M}}\right) \right]
+ {c_{2}}X+\frac{3{c}_{3}X}{\bar{M}^{2}}\left( \frac{\phi ^{\prime } \bar{M}%
^{\prime}}{\bar{M}}-\phi ^{\prime \prime }\right)  \notag \\
&&-\frac{2{c}_{4}X}{\bar{M}^{2}}\left[ \frac{X\bar{N}^{\prime \prime }} {%
\bar{N}}-\frac{5X\bar{M}^{\prime }}{\bar{M}r}+\frac{4\phi ^{\prime }
\phi^{\prime \prime}}{\bar{M}^{2}r}+\frac{\bar{N}^{\prime }} {\bar{N}}\left( 
\frac{X}{r}- \frac{5X\bar{M}^{\prime }}{\bar{M}}+\frac{4\phi ^{\prime }\phi
^{\prime \prime }}{\bar{M}^{2}}\right) \right] =0\,,  \label{GLeq3}
\end{eqnarray}
which are of second order. The equations of motion for the odd-mode
perturbations are also of second order. The stability conditions (\ref{cond2}%
) translate to 
\begin{equation}
c_{4}\left( \frac{X}{M_{\mathrm{pl}}}\right) ^{2}>-\frac{1}{2}\,.
\label{stacova}
\end{equation}
The radial and tangential sound speeds read
\begin{equation}
c_{r}^{2}=1+2c_4 \left( \frac{X}{M_{\mathrm{pl}}} \right)^2\,, \qquad
c_{\Omega }^{2}=1\,,  \label{crcova}
\end{equation}
respectively.

\subsubsection{Covariant Galileons}

Higher-order derivatives present for the covariantized Galileon in a general
curved space-time can be eliminated by including a non-minimally coupled
gravitational contribution to the Lagrangian \cite{Deffayet1,Deffayet2}. The
Galileon model with second-order equations of motion is dubbed
\textquotedblleft covariant Galileon\textquotedblright . This is a sub-class
of the Horndeski Lagrangians (\ref{G2})-(\ref{G4}) with the choice 
\begin{equation}
G_{2}=\hat{c}_{2}X\,,\qquad G_{3}=\hat{c}_{3}X\,,\qquad G_{4}
=\frac{M_{\mathrm{pl}}^{2}}{2}+\hat{c}_{4}X^{2}\,,  \label{covGalcounter}
\end{equation}
where $\hat{c}_{2,3,4}$ are constants.

{}From Eqs.~(\ref{GLPVeom00})-(\ref{GLPVeom22}) the background equations of
motion are given by 
\begin{eqnarray}
&&\frac{M_{\mathrm{pl}}^{2}}{r}\left( \frac{1}{r}-\frac{1}{\bar{M}^{2}r}+ 
\frac{2\bar{M}^{\prime }}{\bar{M}^{3}}\right) +\hat{c}_{2}X+\frac{2\hat{c}
_{3}X}{\bar{M}^{2}}\left( \frac{\phi ^{\prime }\bar{M}^{\prime }}{\bar{M}}
-\phi ^{\prime \prime }\right) +\frac{6\hat{c}_{4}X}{\bar{M}^{2}r}\left( 
\frac{\bar{M}^{2}X}{3r}+\frac{X}{r}-\frac{10X\bar{M}^{\prime }}{\bar{M}}+ 
\frac{8\phi ^{\prime }\phi ^{\prime \prime }}{\bar{M}^{2}}\right) =0\,, 
\notag \\
&&\frac{M_{\mathrm{pl}}^{2}}{r}\left( \frac{1}{r}-\frac{1}{\bar{M}^{2}r}- 
\frac{2\bar{N}^{\prime }}{\bar{M}^{2}\bar{N}}\right) -\hat{c}_{2}X-\frac{2 
\hat{c}_{3}X\phi ^{\prime }}{\bar{M}^{2}}\left( \frac{2}{r}+\frac{\bar{N}
^{\prime }}{\bar{N}}\right) +\frac{30\hat{c}_{4}X^{2}}{\bar{M}^{2}r}\left( - 
\frac{\bar{M}^{2}}{5r}+\frac{1}{r}+\frac{2\bar{N}^{\prime }}{\bar{N}}\right)
=0\,,  \notag \\
&&\frac{M_{\mathrm{pl}}^{2}}{\bar{M}^{2}}\left[ \frac{\bar{M}^{\prime }} {%
\bar{M}r}-\frac{\bar{N}^{\prime \prime }}{\bar{N}}-\frac{\bar{N}^{\prime }} {%
\bar{N}}\left( \frac{1}{r}-\frac{\bar{M}^{\prime }}{\bar{M}}\right) \right]
+ \hat{c}_{2}X+\frac{2\hat{c}_{3}X}{\bar{M}^{2}}\left( \frac{\phi ^{\prime } 
\bar{M}^{\prime }}{\bar{M}}-\phi ^{\prime \prime }\right)  \notag \\
&&+\frac{6\hat{c}_{4}X}{\bar{M}^{2}}\left[ \frac{X\bar{N}^{\prime \prime }} {%
\bar{N}}-\frac{5X\bar{M}^{\prime }}{\bar{M}r}+\frac{4\phi ^{\prime }\phi
^{\prime \prime }}{\bar{M}^{2}r}+\frac{\bar{N}^{\prime }}{\bar{N}}\left( 
\frac{X}{r}-\frac{5X\bar{M}^{\prime }}{\bar{M}}+\frac{4\phi ^{\prime }\phi
^{\prime \prime }}{\bar{M}^{2}}\right) \right] =0\,.
\end{eqnarray}
Compared to the covariantized Galileon, the difference arises from the $B_4$%
-dependent terms in Eqs.~(\ref{GLPVeom00}) and (\ref{GLPVeom11}). The
stability conditions (\ref{cond}) translate to 
\begin{equation}
-\frac{1}{2}<\hat{c}_{4}\left( \frac{X}{M_{\mathrm{pl}}}\right)^{2} <\frac{1%
}{6}\,,
\end{equation}
which is different from Eq.~(\ref{stacova}). The radial and tangential
speeds of sound are given, respectively, by 
\begin{equation}
c_{r}^{2}=\frac{M_{\mathrm{pl}}^{2}-6\hat{c}_{4}X^{2}}{M_{\mathrm{pl}}^{2} +2%
\hat{c}_{4}X^{2}}\,,\qquad c_{\Omega }^{2}=1\,,
\end{equation}
where $c_r^2$ differs from Eq.~(\ref{crcova}).

We have shown that the background and perturbation equations of motion for
both the covariantized Galileon (\ref{covGal}) and the covariant Galileon (%
\ref{covGalcounter}) are of second order on the spherically symmetric
background. Their perturbations propagate identically along the spheres, but
with different propagation speeds in the radial direction.

\section{Concluding Remarks}

\label{conclusec}

We have studied the perturbations about a spherically symmetric and static
background in the framework of the EFT of modified gravity. In this
approach, the gravitational action is expressed in terms of scalar variables
constructed from the canonical variables arising in the
Arnowitt-Deser-Misner (ADM) decomposition of space-time. An additional
scalar field can be included in the gravitational sector at the price of a
partial gauge-fixing (unitary gauge), incorporating it into an explicit
dependence of the dynamics of the coordinate and the lapse associated with
the constant scalar-field hypersurfaces.

Since spherical symmetry selects a preferred radial direction besides the
time direction, we employed a more intricate 2+1+1 decomposition, worked out
earlier for arbitrary dimensions \cite{s+1+1a,s+1+1b}. Due to the double
foliation, there are two sets of extrinsic curvatures in the formalism. Some
of them are related to temporal derivatives $\left( K_{ab}\,,\,\mathcal{K}%
_{a}\,,\,\mathcal{K}\right) $, the others to radial derivatives $\left(
L_{ab}\,,\,\mathcal{L}_{a}\,,\,\mathcal{L}\right)$. We have started from a
general action that depends on scalars formed from these quantities, the
metric variables of the constant time hypersurfaces ($h_{ab}$, $M_a$, $M$)
and the lapse $N$.

We choose the gauge $\mathcal{N}=0$ to ensure the perpendicularity of the
foliations on the spherical symmetric space-time. Then, the dynamics of the
radial and temporal components proceeds in a hypersurface-orthogonal manner
without vorticities. By this gauge choice, it is possible to avoid an
unnecessary increase in the number of variables associated with
vorticity-type quantities. A second gauge fixing is the radial unitary gauge 
$\phi=\phi(r)$, which switches off the perturbations of the scalar field 
($\delta \phi=0$). In this case, the scalar field is absorbed in the
gravitational sector (into the radial lapse $M$) and an explicit radial
dependence of the action.

The gravitational action (\ref{action}) incorporates a general system of a
single scalar degree of freedom. Despite the relatively large number of
scalar variables, variation of the action gives rise to three independent
equations of motion at the background level. They are derived by the changes
in the lapse $\delta N$, in the radial lapse $\delta M$, and in the scalar
curvature on the sphere $\delta \mathcal{R}$, respectively. 
Equations (\ref{bg00})-(\ref{bg22}) represent the most generic set of equations of motion
in modified gravity theories on the spherically symmetric and static
background.

The Horndeski theory and their recently suggested GLPV 
generalizations \cite{Gleyzes} involve a single scalar degree of freedom 
beside the metric. They
represent the most general class of theories with second-order equations of
motion and the extended class that allows for higher-order derivatives in
generic space-time, respectively. We have expressed both Lagrangians in
terms of the 2+1+1 variables, proving that they belong to the class of the
EFT of modified gravity studied in this paper. We also derived the
background equations of motion explicitly for both under spherical symmetry
and staticity. Under these symmetries the GLPV background is also second
order, as in the case of the Horndeski theory.

Variation of the action up to second order leads to the linear perturbation
equations of motion, with the even and odd modes decoupled. In this paper we
focused on the analysis for the odd-parity mode of perturbations. The
originally fourth-order differential equations were reduced to second order
by employing a multipolar expansion into spherical harmonics. 
We derived the second-order Lagrangian density for odd-mode 
perturbations of the form (\ref{quad5}). 
Under the condition (\ref{cond1}), which is satisfied for both Horndeski and 
GLPV theories, the Lagrangian density is expressed solely by 
a dynamical scalar variable $Z_l$ and its derivatives.
We established extremely simple conditions for avoiding ghosts and Laplacian
instabilities. The propagation speed of odd-mode perturbations depends on
the direction of propagation. More specifically, the radial sound speed and
the sound speed along the spheres are different, generalizing the
corresponding result established for the Horndeski theory \cite{suyama}.

As applications of our general stability analysis, we have i) confirmed the
corresponding results for the Horndeski theory, ii) obtained the stability
conditions for the recently proposed GLPV theory, iii) derived and compared
both the tangential and the radial speeds of sound for two types of Galileon
theories: \textquotedblleft covariantized Galileon\textquotedblright\
(derived by replacing coordinate derivatives with covariant derivatives in
the original Galileon model) and \textquotedblleft covariant
Galileon\textquotedblright\ with second-order dynamics in general space-time
(obtained by adding a new term to eliminate higher-order derivatives).
Although the background equations of motion are similar in the two Galileon
theories, the stability conditions associated with the radial propagation
speed $c_{r}$ are different. This can be traced back to the terms $B_{4}$
and $B_{5}$ appearing in the Lagrangians (\ref{exhornLa4}) and 
(\ref{exhorn}) being different in these two theories. 
In the Horndeski theory $B_{4}$ and 
$B_{5}$ are related to the other terms $A_{4}$ and $A_{5}$ according to
Eqs.~(\ref{H1}) and (\ref{H2}), however in general no such restriction
appears in the GLPV theory.

Recently, the cosmology based on the two Galileon theories was studied in
Ref.~\cite{Katsu} on the flat Friedmann-Lema\^{\i}tre-Robertson-Walker
background. It was shown that the propagation speeds of the field $\phi $
for covariant and covariantized Galileons are different due to the different
values of $B_{4}$ and $B_{5}$ in the two theories. On the isotropic
cosmological background, the equations of motion for linear perturbations
also remain of second order. In spite of the possible presence of
derivatives higher than second order on general backgrounds, the GLPV theory
remains healthy on both the static spherically symmetric and the isotropic
cosmological backgrounds.

It is possible to extend our work to several interesting directions. First,
the background equations of motion (\ref{bg00})-(\ref{bg22}) can be
generally applied to the discussion of the screening mechanism of the fifth
force mediated by the scalar field $\phi$. Second, the analysis of
even-parity perturbations, which is much more involved than that of
odd-parity modes, will be useful to discuss the full stability of the EFT of
modified gravity on the spherically symmetric and static background. Third,
the construction of theoretically consistent dark energy models in the
framework of the GLPV theory will be also intriguing. We leave these issues
for future works.

\section*{Acknowledgements}

We are grateful to Riccardo Penco and Federico Piazza for interactions in
the early stages of this project. R.\ K.\ and S.\ T.\ were supported by the
Scientific Research Fund of the JSPS (Nos.~24$\cdot $6770 and 24540286).
L. Á. G. was supported by the long-term Invitation 
Fellowship Program no. L13519 of the Japan Society 
for the Promotion of Science (JSPS). 


\appendix
\section{Equations of motion in the Horndeski and GLPV theories on the
spherically symmetric and static background}
\label{sec:eomhorn} 

In this Appendix we present the background equations of motion for the
spherically symmetric and static GLPV theory (including the Horndeski
theory). Substituting the Lagrangians (\ref{exhornL2})-(\ref{exhornL4}) into
Eqs.~(\ref{bg00})-(\ref{bg22}), it follows that 
\begin{eqnarray}
&&A_{2}-\frac{\phi ^{\prime }A_{3\phi }+\bar{X}^{\prime }A_{3X}}{\bar{M}} +%
\frac{2A_{4}}{\bar{M}^{2}r}\left( \frac{1}{r}-\frac{2\bar{M}^{\prime }} {%
\bar{M}}\right) +\frac{4\left( \phi ^{\prime }A_{4\phi }+\bar{X}^{\prime
}A_{4X}\right) }{\bar{M}^{2}r}+\frac{2B_{4}}{r^{2}}=0\,,  \label{GLPVeom00}
\\
&&A_{2}-2\bar{X}A_{2X}-\frac{2\bar{X}A_{3X}}{\bar{M}}\left( \frac{2}{r}+%
\frac{\bar{N}^{\prime }}{\bar{N}}\right) +\frac{2\left( A_{4}+2\bar{X}%
A_{4X}\right) }{\bar{M}^{2}r}\left( \frac{1}{r} +\frac{2\bar{N}^{\prime }}{%
\bar{N}}\right) +\frac{2\left( B_{4}-2\bar{X}B_{4X}\right) }{r^{2}}=0\,,
\label{GLPVeom11} \\
&&A_{2}-\frac{\phi ^{\prime }A_{3\phi }+\bar{X}^{\prime } A_{3X}}{\bar{M}}-%
\frac{2A_{4}}{\bar{M}^{2}}\left[ \frac{\bar{N}^{\prime \prime }}{\bar{N}}+ 
\frac{\bar{M}^{\prime }}{\bar{M}r}-\frac{\bar{N}^{\prime }} {\bar{N}}\left( 
\frac{1}{r}-\frac{\bar{M}^{\prime }}{\bar{M}}\right) \right] +\frac{2\left(
\phi ^{\prime }A_{4\phi }+\bar{X}^{\prime } A_{4X}\right) }{\bar{M}^{2}}
\left( \frac{1}{r}+\frac{\bar{N}^{\prime }}{\bar{N}}\right) =0\,,
\label{GLPVeom22}
\end{eqnarray}
where $\bar{X}$ represents the background value of the kinetic term $X$,
i.e., $\bar{X}=\phi ^{\prime 2}/\bar{M}^{2}$. The last terms on the lhs of
Eqs.~(\ref{GLPVeom00})-(\ref{GLPVeom11}), which include $B_{4}$ and its
derivative with respect to $X$, originate from the non-vanishing
two-dimensional scalar curvature $\mathcal{R}$ on the spherically symmetric
and static background. In the Horndeski theory $B_{4}$ is entirely
determined by $A_{4}$ and $X$, while in the GLPV theory it is not. Hence the
equations of motion for the GLPV theory generally differ form those for the
Horndeski theory\footnote{%
On the flat isotropic cosmological background the scalar curvature of the
constant time hypersurfaces identically vanishes. We verified that no $B_{4}$
terms appear in the background equations of motion of the GLPV theory, which
then coincide with those of the Horndeski theory at the background level.}.

Under the condition (\ref{H1}) and by redefining the functions $A_{2}$, $%
A_{3}$ and $B_{4}$ in terms of the new functions $G_{2}$, $F_{3}$ and $G_{4}$
as follows 
\begin{equation}
A_{2}=G_{2}-F_{3\phi }X\,,\qquad A_{3}=2X^{3/2}F_{3X}+2\sqrt{X}G_{4\phi
}\,,\qquad B_{4}=G_{4}\,,
\end{equation}
the sum of the Lagrangians $L_{2,3,4}^{\mathrm{GLPV}}$ manifestly reduces to
that of $L_{2,3,4}^{\mathrm{H}}$. Applying the same condition and
redefinitions to the equations of motion (\ref{GLPVeom00})-(\ref{GLPVeom22}%
), we obtain those for the Horndeski theory. In order to compare them with
the equations of motion derived in Ref.~\cite{Kase} by a method entirely
intrinsic to the Horndeski theory, we further need the conversion in the
notations ($\bar{N}$, $\bar{M}$, $X$, $G_{3}$)$\rightarrow $($e^{\Psi (r)}$, 
$e^{\Phi (r)}$, $-2X$, $-G_{3}$), after which a full agreement is reached.


\end{document}